\documentclass[useAMS,usenatbib,letterpaper]{mn2e}
\usepackage{graphicx,amsmath,color,amssymb}
\usepackage[pdftitle={Moving mesh cosmology: properties of neutral hydrogen in absorption},pdfauthor={Simeon Bird}]{hyperref}
\usepackage[all]{hypcap}
\usepackage{subfigure}
\newcommand{\adsurl}[1]{\href{#1}{ADS}}

\newlength{\narrowFigurewidth}
\newlength{\Figurewidth}
\newlength{\wideFigurewidth}
\newcommand{\etal}
  {et al.}

\newcommand{\Lya}{Lyman-$\alpha\;$}
\newcommand{\Msun}{\, h^{-1} M_\odot}
\newcommand{\NHunit}{atoms cm$^{-2}$}
\newcommand{\sLLS}{\sigma_\mathrm{LLS}}

\newcommand{\Mpch}{\, h^{-1} \mathrm{Mpc}}
\newcommand{\kpch}{\, h^{-1}\mathrm{kpc}}

\newcommand{\gadget}{{\small GADGET\,}}
\newcommand{\arepo}{{\small AREPO\,}}

\newcommand{\NHI}{N_\mathrm{HI}}
\newcommand{\sigmaDLA}{\sigma_\mathrm{DLA}}

\newcommand{\edit}[1]{#1}

\title[Moving mesh cosmology: neutral hydrogen absorption]{Moving mesh
  cosmology: properties of neutral hydrogen in absorption}

\author
  [S. Bird \etal]
  {Simeon Bird$^{1}$\thanks{E-mail: spb@ias.edu},
  Mark Vogelsberger$^{2}$, Debora Sijacki$^{2}$, Matias Zaldarriaga$^{1}$, 
\newauthor Volker Springel$^{3,4}$, Lars Hernquist$^{2}$ \vspace{4mm}\\
$^1$Institute for Advanced Study, 1 Einstein Drive, Princeton, NJ, 08540, USA\\
$^2$Harvard-Smithsonian Center for Astrophysics, 60 Garden Street, Cambridge, MA 02138, USA \\
$^3$Heidelberg Institute for Theoretical Studies, Schloss-Wolfsbrunnenweg 35, 69118 Heidelberg, Germany \\
$^4$Zentrum f\"{u}r Astronomie der Universit\"{a}t Heidelberg, ARI, M\"{o}nchhofstr. 12-14, 69120 Heidelberg, Germany \\
}

\begin{document}

\pagenumbering{alph}
\date{}

\maketitle
%The \pageref{lastpage} is broken
\pagerange{\pageref{firstpage}--\pageref{lastpage}} \pubyear{2012}

\pagenumbering{arabic}
\label{firstpage}

\begin{abstract}
  We examine the distribution of neutral hydrogen in cosmological
  simulations carried out with the new moving-mesh code \arepo and
  compare it with the corresponding \gadget~simulations based on the smoothed
  particle hydrodynamics (SPH) technique. The two codes use identical
  gravity solvers and baryonic physics implementations, but very
  different methods for solving the Euler equations, allowing us to
  assess how numerical effects associated with the hydro-solver impact
  the results of simulations. Here we focus on an analysis of the neutral
  gas, as detected in quasar absorption lines. We find that the high
  column density regime probed by Damped \Lya (DLA) and Lyman Limit
  Systems (LLS) exhibits significant differences between the
  codes. \gadget~produces spurious artefacts in large halos in the
  form of gaseous clumps, boosting the LLS cross-section.
  Furthermore, it forms halos with denser central baryonic 
  cores than \arepo, which leads to a substantially greater DLA cross-section from
  smaller halos.  \arepo~thus produces a significantly lower cumulative
  abundance of DLAs, which is intriguingly in much closer agreement with observations. 
  \edit{The column density function, however, is not altered 
  enough to significantly reduce the discrepancy with the observed value.}
  For the low column density gas probed by the \Lya forest, the codes
  differ only at the level of a few percent, suggesting that this
  regime is quite well described by both methods, a fact that is
  reassuring for the many \Lya studies carried out with SPH thus
  far. While the residual differences are smaller than the errors on
  current \Lya forest data, we note that this will likely change for future
  precision experiments.
\end{abstract}

 
\section{Introduction}
\label{sec:intro}

Absorption features in quasar spectra offer a unique view of structure
formation at redshifts  $z=2-4$. \Lya absorption of neutral hydrogen 
directly tracks the distribution of gas during the initial stages of galaxy formation.
Damped \Lya Systems (DLAs) have a neutral
hydrogen column density $\NHI > 10^{20.3}\,{\rm cm}^{-2}$ \citep{Wolfe:1986}, 
and thus can be observed through natural broadening of the \Lya line.
\edit{DLAs are thought to be high redshift proto-galaxies, sufficiently dense that their interiors
are shielded from the ionising effect of the diffuse radiation
background \citep{Katz:1996a, Nagamine:2010}. Thus at $z=2-4$ they are understood to be
reservoirs containing most of the neutral hydrogen in the Universe \citep{Gardner:1997}, 
corresponding either to large discs \citep{Prochaska:1997, Maller:2001} or 
irregular protogalactic clumps \citep{Haehnelt:1998, Okoshi:2005}.
Recent kinematic data may prefer large discs \citep{Barnes:2009, Hong:2010}.
A wide range of quasar surveys have gradually increased the 
available sample of high-redshift DLAs \citep[e.g.][]{Wolfe:1995, Storrie:2000, Peroux:2005, Prochaska:2005, Noterdaeme:2009}, 
low-redshift DLAs \citep{Rao:2000, Prochaska:2001,Chen:2003}
and their lower column density cousins, Lyman Limit Systems (LLS) \citep{Peroux:2001,OMeara:2007, Prochaska:2009}. 
LLS are defined to be absorbers with $10^{17}\,{\rm cm}^{-2} < \NHI < 10^{20.3}\,{\rm cm}^{-2}$.
They have been connected to lower density analogues of DLAs \citep{Gardner:2001} or 
filamentary structures on the outskirts of proto-galaxies \citep{Fumagalli:2011, Faucher:2011}, 
and are affected by shielding \citep{McQuinn:2011a}.}
At lower column densities, we see the \Lya forest; a complex region of
overlapping \Lya lines. The \Lya forest is a probe of the matter
distribution in the low-density intergalactic medium
\citep{Hernquist:1996, Miralda:1996, Dave:1999, Gnedin:1998, Croft:1998}.
It has been used to constrain the initial
conditions of the Universe \citep[e.g.][]{Seljak:2005, Viel:2005, 
  Viel:2009, Bird:2011}, the processes that govern the thermal
state of the gas \citep[e.g.][]{Faucher:2008b,Bolton:2008,Lidz:2010}
and indirectly helium reionisation \citep{Faucher:2008a, Becker:2010}.

The \Lya forest is produced by diffuse absorbers $\sim 100 \kpch$ 
across, collapsing under gravity. 
DLAs, on the other hand, are produced in smaller, denser systems
and are strongly influenced by gas physics. 
In both cases, obtaining accurate quantitative results for their
properties requires following non-linear gravitational collapse with 
N-body simulation techniques and a method for solving the inviscid 
Euler equations for the cosmic gas.
Historically two main approaches have been used when tackling this
problem through direct simulations: grid-based codes that discretise the
gas on an adaptively refined mesh \citep{Berger:1989, Teyssier:2002,
  OShea:2004} and particle-based codes that use the smoothed particle
hydrodynamics (SPH) technique \citep{Lucy:1977, Gingold:1977, Monaghan:1992,
  Monaghan:2005}. The former solves the equations of motion
for a fluid on a stationary grid of cells in an Eulerian fashion. The
latter is a pseudo-Lagrangian technique, where the fluid is split into
a number of discrete mass elements, which are assumed to be
indivisible and are then followed as particles. Smoothed fluid quantities
are constructed from the particles through kernel interpolation.

Each approach has its own advantages and disadvantages, making it
non-trivial to judge their relative accuracy for different
applications. For example, while SPH codes have excellent conservation
properties, they suffer from large gradient errors and accuracy
problems in resolving fluid instabilities \citep{Agertz:2007},
something that can make them fail dramatically in some idealised fluid
dynamics problems \citep{Sijacki:2012}. Eulerian mesh codes on the
other hand offer high accuracy for capturing shocks, but many have incorporated relatively
inaccurate gravity solvers \citep[see, e.g.][]{Oshea:2005}. 
Furthermore, their truncation error depends on the absolute velocity of the gas, 
which is problematic for the often highly supersonic flows encountered
in cosmological simulations.

\cite{Springel:2010} proposed a new technique which aims to combine
the strengths of both approaches. In this method, space is tessellated
with a moving mesh; cells advect with the fluid flow, and as a result
each cell contains approximately the same gas mass. Thus the 
resolution automatically increases in areas of higher density, just as
in Lagrangian methods like SPH. A moving mesh method can resolve shocks 
equally as well as grid methods, since in both cases the Euler equations are solved
with a high-accuracy finite volume method, but without the advection
errors present in Eulerian codes. The \arepo\ code is an
implementation of this technique and has been shown to perform well
in many idealised simulations of fluid dynamical problems
\citep{Springel:2010, Sijacki:2012}.

We compare results obtained with \arepo~to corresponding
simulations with the well-tested SPH code \gadget, last described in
\citet{Springel:2005}.  Although we know that SPH cannot accurately
describe fluid instabilities and mixing processes in its widely
employed `standard' form, it is not obvious a priori to what extent
these inaccuracies are important for different aspects of structure
formation, or how they affect our interpretation of observational
data.  This paper therefore examines the important issue of how
predictions for the neutral hydrogen distribution in the Universe
depend on the employed numerical technique. 

Our study is part of a
series which compares simulations run using \arepo~and \gadget~with
identical initial conditions, gravity solver, and baryonic physics
parameters. They thus differ only in the approach to hydrodynamics,
offering an unrivalled opportunity to isolate the effects of numerical
uncertainties. In previous work, \cite{Vogelsberger:2012} examined the
global properties of baryons and halos, as well as performing
convergence and numerical tests on \arepo. \cite{Sijacki:2012} studied
a number of idealised fluid problems to clarify the origins of the
observed differences. \cite{Keres:2012} looked at the effect on
the resulting galaxy properties, and \cite{Torrey:2012} focussed on the
structure of galactic discs.  Here we shall look at how the use
of a moving-mesh technique affects the properties of DLAs, LLS and the \Lya
forest.  Neutral hydrogen is particularly relevant as it is a comparatively clean
probe of the hydrodynamics. 
\edit{We include star formation, but neglect strong feedback from outflows.
\cite{Scannapieco:2012} examined the effect of different feedback models and 
hydrodynamic solvers on a single collapsed object at $z=0$, and found that it 
was more strongly affected by the choice of feedback model.
Omitting strong feedback thus helps to avoid the possibility that
necessary differences in the feedback implementation may affect our results 
for larger halo samples.} Also, rather than a full radiative transfer model, we shall 
use a simple density cut-off to account for self-shielding of the
gas.  While less sophisticated than many previous works, this simple
approach allows us to focus on the effect of the hydro solver and
helps to connect our intuition from idealised tests with observations.

We build on a large literature of simulated DLAs:
\citet{Nagamine:2004a, Nagamine:2004b} used SPH simulations with
\gadget\ to examine their abundance and metallicity. They also
examined the effects of galactic winds on DLAs, looked at further by
\cite{Nagamine:2007} and \cite{Tescari:2009}.  \cite{Pontzen:2008} attempted to
reproduce many observed properties of DLAs with a complete simulation
framework incorporating a simple model of radiative transfer and
supernova feedback. Radiative transfer was also examined in more
detail by \cite{Yajima:2011}. \edit{\cite{Erkal:2012} and 
\cite{Altay:2011} looked at the effects of molecular hydrogen.}
\cite{Cen:2012} and \cite{Fumagalli:2011} both used Eulerian grid codes with adaptive mesh
refinement to study DLAs and LLS.  They included feedback from star
formation and metal cooling, which we neglect, but their AMR-based
simulations were unable to resolve halos less massive than $10^{10}
\Msun$.

This paper is structured as follows. In Section~\ref{sec:methods}, we
discuss our methods in more detail.  We then present our results in
Section \ref{sec:results} and conclude with a summary of our findings
in Section~\ref{sec:conclusions}.

\section{Methods}
\label{sec:methods}

\subsection{Numerical codes and simulation set}

Both \arepo\ and \gadget\ compute gravitational interactions using the TreePM 
approach, as described in \cite{Springel:2005}. Radiative cooling is 
implemented, following \cite{Katz:1996}, using a rate network, including line cooling, free-free emission 
and inverse Compton cooling off the cosmic microwave background. We assume ionisation equilibrium 
and optically thin gas during the simulation, accounting for gas self-shielding 
by post-processing the simulation outputs, as described in Section~\ref{sec:nhifrac}.
The history of the ultraviolet background (UVB) follows the 
estimates of \citet{Faucher:2009}. 

Star formation is implemented with the effective two-phase model of
\cite{Springel:2003}. We use the same parameters as in that paper,
giving a threshold density for star formation of $N_h = 0.13\,{\rm
  cm}^{-3}$.  The star formation time-scale is assumed to scale with
density as $t_{\star}\propto \rho^{-0.5}$, normalized to $2.1$ Gyr at
the threshold density in order to match the local relation between the 
gas surface density and star formation rate of galaxies
\citep{Kennicutt:1998}. The energy associated with supernovae heats
the multi-phase medium and thereby regulates star formation, but in
the model implementation used here this feedback is not strong
enough to drive significant outflows. We note that
\cite{McDonald:2005, Cen:2005} found that galactic winds have little
effect on the \Lya forest, but \cite{Nagamine:2007, Tescari:2009}
showed that they do affect the DLA cross-section and column density
function. Since this work is primarily aiming to compare the relative
performance of two different codes, the absence of strong winds is not
a problem. Indeed, it helps to avoid more complicated gas motions 
which could obscure our results.

We use the simulations first described in \cite{Vogelsberger:2012}. 
The initial conditions were generated at $z=99$, using the power spectrum fit of 
\cite{Eisenstein:1999} with cosmological parameters $\Omega_m = 0.27$,
$\Omega_{\Lambda} =0.73$, $\Omega_b = 0.045$, $\sigma_8 = 0.8$, $n_s = 0.95$
and $H_0 =70\, \rm km s^{-1} Mpc^{-1}$ ($h=0.7$), consistent with the latest 
WMAP results \citep{WMAP}. The box size is $20 \Mpch$.

Most of our results are from the highest resolution simulation of
\cite{Vogelsberger:2012}.  This has $512^3$ dark matter particles,
with a particle mass of $3.722 \times 10^6 \Msun$, and a comoving gravitational
softening length of $1 \kpch$ ($1/40$ of the mean inter-particle
spacing).  Each simulation was initialised with $512^3$ gas elements,
with an initial mass of $7.44 \times 10^5 \Msun$. However, 
the number and mass of gas elements are both altered over time by star
formation, and, in \arepo, mass fluxes, refinement and
de-refinement of grid cells. Further details on the refinement
implementation used in \arepo\ may be found in
\cite{Vogelsberger:2012}.  We examined three output times
corresponding to redshifts $z=4$, $3$ and $2$.  We emphasize that all
the simulations used the same realisation for their initial conditions
as well as the same parameters for the sub-grid physics model,
allowing a comparison on an object-by-object basis.

We refer the reader to \cite{Springel:2010} for further details of the
moving mesh implemented in \arepo, to \cite{Springel:2005} for details
of the gravity computation, to \cite{Springel:2002} for the SPH
implementation in \gadget, and to \cite{Vogelsberger:2012} for a full
account of the parameters of the simulations. In the rest of
this Section, we focus on those aspects of the analysis specifically
concerned with DLAs.

\subsection{Gas self-shielding}
\label{sec:nhifrac}

The gas in DLAs is self-shielded, with a neutral fraction close to unity. 
A full treatment requires radiative transfer and has been studied in, e.g.,
\cite{Pontzen:2008, Altay:2011, Yajima:2011, Cen:2012}. 
\cite{Yajima:2011}, using outputs from an SPH simulation post-processed 
with radiative transfer, suggested modelling the transition 
to self-shielded gas with a step function. 
We considered this, but found that it caused an artificial kink in the 
column density function, especially prominent for \arepo, although it did not affect any 
of our other results. Instead we fit the neutral fraction as a function of density 
to the results of their preferred simulation. Gas is assumed to transition between 
equilibrium with the UVB for $ \rho < \rho_\mathrm{U}$ and complete self-shielding 
for $\rho > \rho_\mathrm{s}$, with the transition region given by
\begin{equation}
 n_\mathrm{HI} = \frac{n_\mathrm{HI}^\mathrm{UVB} ( \rho_\mathrm{s} - \rho )^p + ( \rho - \rho_\mathrm{U} )^p}{( \rho_\mathrm{s} - \rho_\mathrm{U} )^p}\,,
\end{equation}
where $n_\mathrm{HI}^\mathrm{UVB}$ is the neutral fraction when in equilibrium with the UVB. 
The best fit values were $p=2.68$, $\rho_\mathrm{U} = 4.53 \times 10^{-3}$ {\rm cm}$^{-3}$ and 
$\rho_\mathrm{s} = 1.52 \times 10^{-2}$ {\rm cm}$^{-3}$. 
Furthermore, we verified that this was consistent with the results of \cite{Altay:2011}, 
and that the DLA abundance is not sensitive to the exact position of the transition. 

\cite{Nagamine:2004a} proposed identifying self-shielding with the
onset of star-formation; i.e., $\rho_\mathrm{ss} = 0.1289\, {\rm
  cm}^{-3}$.  This is an order of magnitude higher than the value of
$\rho_\mathrm{ss}$ suggested by radiative transfer and produces a
slight, probably unphysical, coupling of the neutral fraction to the
star formation rate.  While we did not use this prescription, we did
check its effect and found that, although the absolute value of many
DLA properties changed (which is to be expected when changing the
density threshold for self-shielding by an order of magnitude), the
differences between \arepo\ and \gadget\ were largely unaffected. This
gives us further confidence that our results are robust to changes in
the self-shielding prescription.

\cite{Cen:2012} and \cite{Altay:2011} included a prescription for the formation of molecular hydrogen. 
\cite{Altay:2011} were motivated by \cite{Blitz:2006}, 
who found that the surface density of molecular hydrogen in local galaxies was strongly correlated
with the hydrostatic pressure. \cite{Altay:2011} assumed the same relation held between molecular hydrogen density 
and pressure in the star-forming phase of the ISM at high redshift. 
We considered such a prescription, but found the only noticeable effect was 
on the column density function for 
$N_\mathrm{HI} > 10^{22}$ \NHunit (as found by \cite{Erkal:2012}), and was not sufficient to bring it into agreement with observations. 
Furthermore, we compared the shape and amplitude of the molecular 
hydrogen column density function predicted by the model at $z=0$ to the observed values of \cite{Zwaan:2006} and 
found that they did not match. We believe both these facts are because 
the lack of feedback in our simulations led to a surplus of over-dense material. 
We therefore decided not to incorporate molecular hydrogen in our 
analysis until our simulations include strong feedback processes.

\subsection{DLA selection and column density}
\label{sec:projection}

We use a halo catalogue generated with the Friends-of-friends (FOF) 
algorithm \citep{Davis:1985} and a linking length of $0.2$ of the 
mean inter-particle spacing. The FOF algorithm is applied only to the 
DM particles, whereas baryonic particles/cells are later assigned to their nearest DM particle and
included in the halo to which the corresponding DM particle belongs.
Self-bound concentrations of mass are identified
within each FOF halo using the {\small SUBFIND} algorithm \citep{Springel:2001},
modified to account correctly for baryons \citep{Dolag:2009}.
{\small SUBFIND} identifies halo substructure by generating an adaptively 
smoothed density field and searching for gravitationally bound over-densities. 
Following a common convention in the literature, we use $M_{200}$ as an estimate of
the halo virial mass. $M_{200}$ is the mass enclosed within a spherical region
of radius $R_{200}$ within which the mean density is $200$ times the
critical density.

We consider only resolved halos with $M > 400 (\Omega_m /
\Omega_b)\,m_{b}$, where $m_b$ is both the SPH gas particle mass and the initial
mass of \arepo\ cells.  For our $2\times 512^3$ simulation, this
implies a halo resolution threshold of $M \ge 2 \times 10^9 \Msun$.
We note that our conclusions are unaffected by the exact placing
of this limit. To avoid confusing overlaps, we remove halos with 
larger neighbours closer than $R_{200}$ from our catalogue.

To find the projected neutral hydrogen column density around each
halo, we consider a grid of size $2 R_{200}$ centred on each halo,
divided into equal-sized cells with side length equal to the
gravitational softening length (for us $\sim 1 \kpch$).  Gas elements
in the halo are projected onto this grid, using an SPH kernel.  In the
case of \arepo, the SPH smoothing length is chosen so that the volume
covered by the SPH kernel is identical to the volume of the moving
mesh cell. We also considered a cloud-in-cell kernel and, like 
\cite{Nagamine:2004a}, found that it made negligible difference to our results.

We calculate the column density along a line of sight by projecting our interpolated density field along a single direction, 
here the $x$ axis. The column density is given by 
\begin{equation}
 N_\mathrm{HI} = \Sigma_x \frac{\rho_\mathrm{HI}(x) }{m_\mathrm{P}} \epsilon (1+z)^2\,,
\label{eq:column_density}
\end{equation}
where $m_\mathrm{P}$ is the proton mass, $\epsilon$ is the side length of a single grid cell and $\rho_\mathrm{HI} (x)$ 
is the average neutral hydrogen density inside the cell. The factor of $(1+z)^2$ enters because $\rho_\mathrm{HI}$ is 
in comoving units and $N_\mathrm{HI}$ is in physical units.  We checked different projection directions and found 
the overall statistical properties of DLAs were unchanged. However, since galaxies in \arepo~are more disc-shaped, 
they tended to look somewhat different when viewed edge-on. 
We also verified that our results are robust to changes in the resolution of the grid used for the map making.

\subsection{Column density function}
\label{sec:dlaobs}

The column density function, $f(N_\mathrm{HI})$, is defined observationally such
that $f(N_\mathrm{HI})\, {\rm d}N_\mathrm{HI}\, {\rm d}X$ is the number of absorbers per
sight-line with column density in the interval $[N_\mathrm{HI}, N_\mathrm{HI} + {\rm d}N_\mathrm{HI}]$.
We identify sight-lines with grid cells and thus count absorbers by computing 
a histogram of the column density on the grid. 
Equating grid cells with sight-lines assumes that two simulated DLAs will 
rarely be found along the same sight-line, which is a good assumption given the small size of our box.
More explicitly, we define the column density function by 
\begin{align}
 f(N) &= \frac{F(N)}{\Delta N}{\Delta X(z)}\,.
\end{align}
$F(N)$ is the fraction of the total number of grid cells in a given column density bin, 
and $\Delta X(z)$ is the absorption distance per sight-line. 
As described by \cite{Bahcall:1969, Nagamine:2004a}, the (dimensionless) absorption distance is given by
\begin{equation}
 X(z) = \int_0^z (1+z')^2  \frac{H_0}{H(z')} {\rm d}z'\,,
\end{equation}
and, for a box of comoving length $\Delta L$, we have 
$\Delta X = (H_0 /c) (1+z)^2 \Delta L$.

\section{Results}
\label{sec:results}

In this section, we present our results for the comparison between \arepo\ and
\gadget. In Sections \ref{sec:haloshape} and \ref{sec:centralden}, we 
look at the effects on large and small halos, respectively.  Then we examine
various statistical properties of the DLAs: Section \ref{sec:sigma_dla}
considers the DLA and LLS cross-sections, Section \ref{sec:dla_abund} the
observed DLA abundance and Section \ref{sec:columnden} the column density
function.  Finally, in Section \ref{sec:lya} we discuss our results for the
\Lya forest.  We emphasize that we have checked explicitly that our results
are unchanged when using simulations with $2\times 256^3$ particles - a factor
of $8$ lower mass resolution - at least for halos above the resolution limit of
the lower resolution simulations.

\subsection{Large halos}
\label{sec:haloshape}

\begin{figure*}
% \centering
\includegraphics[width=0.45\textwidth]{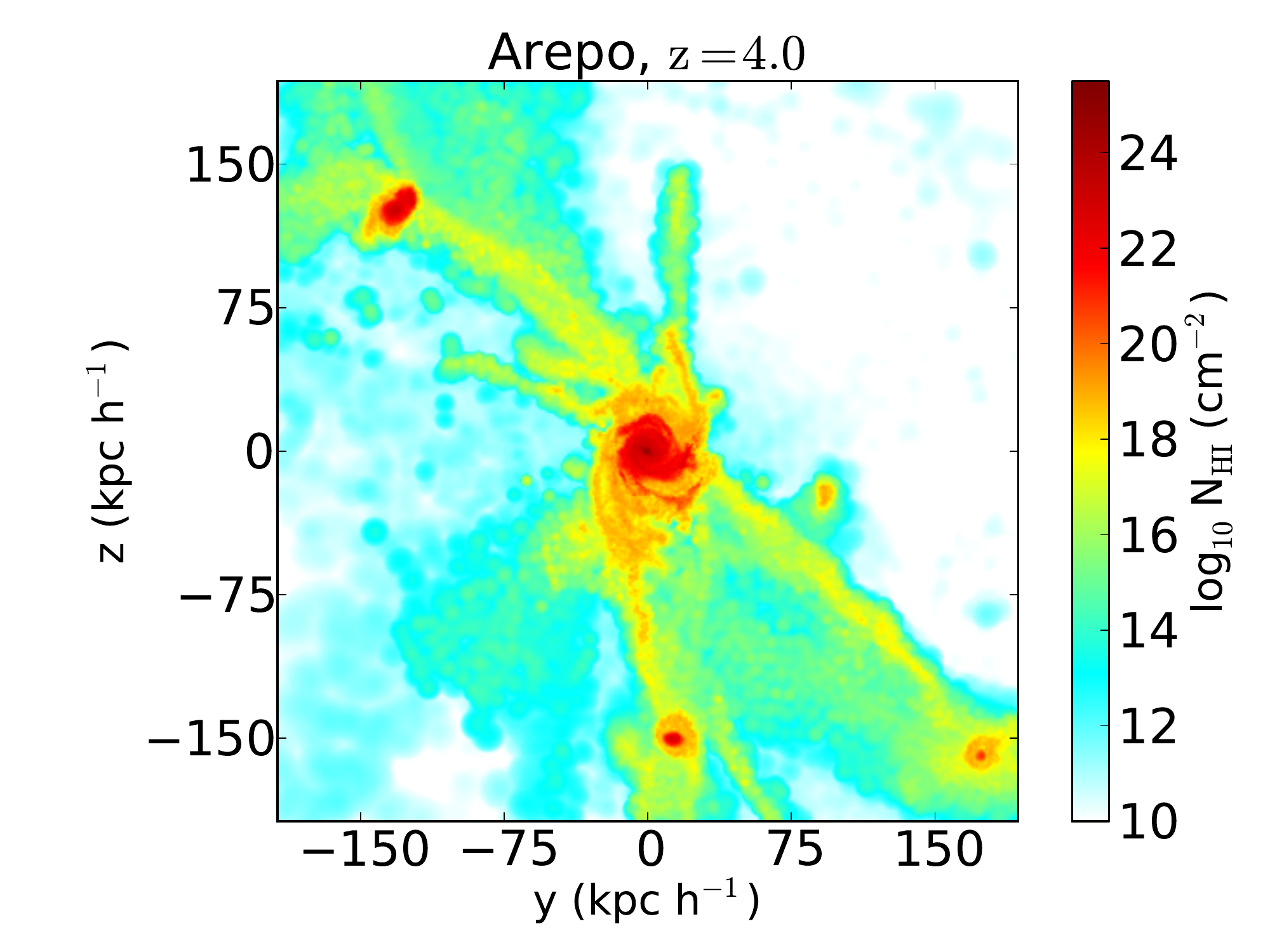}
\includegraphics[width=0.45\textwidth]{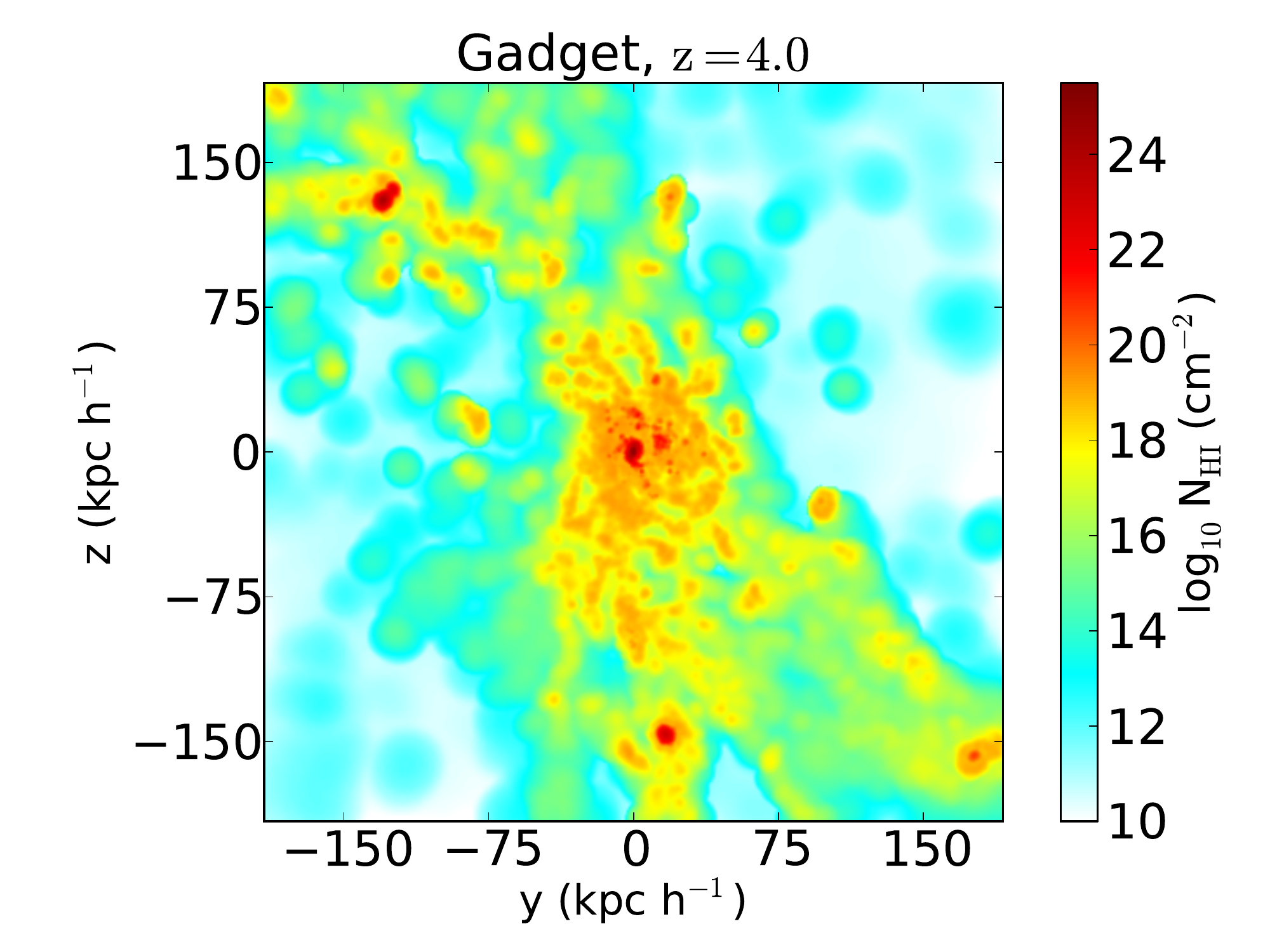} \\
\includegraphics[width=0.45\textwidth]{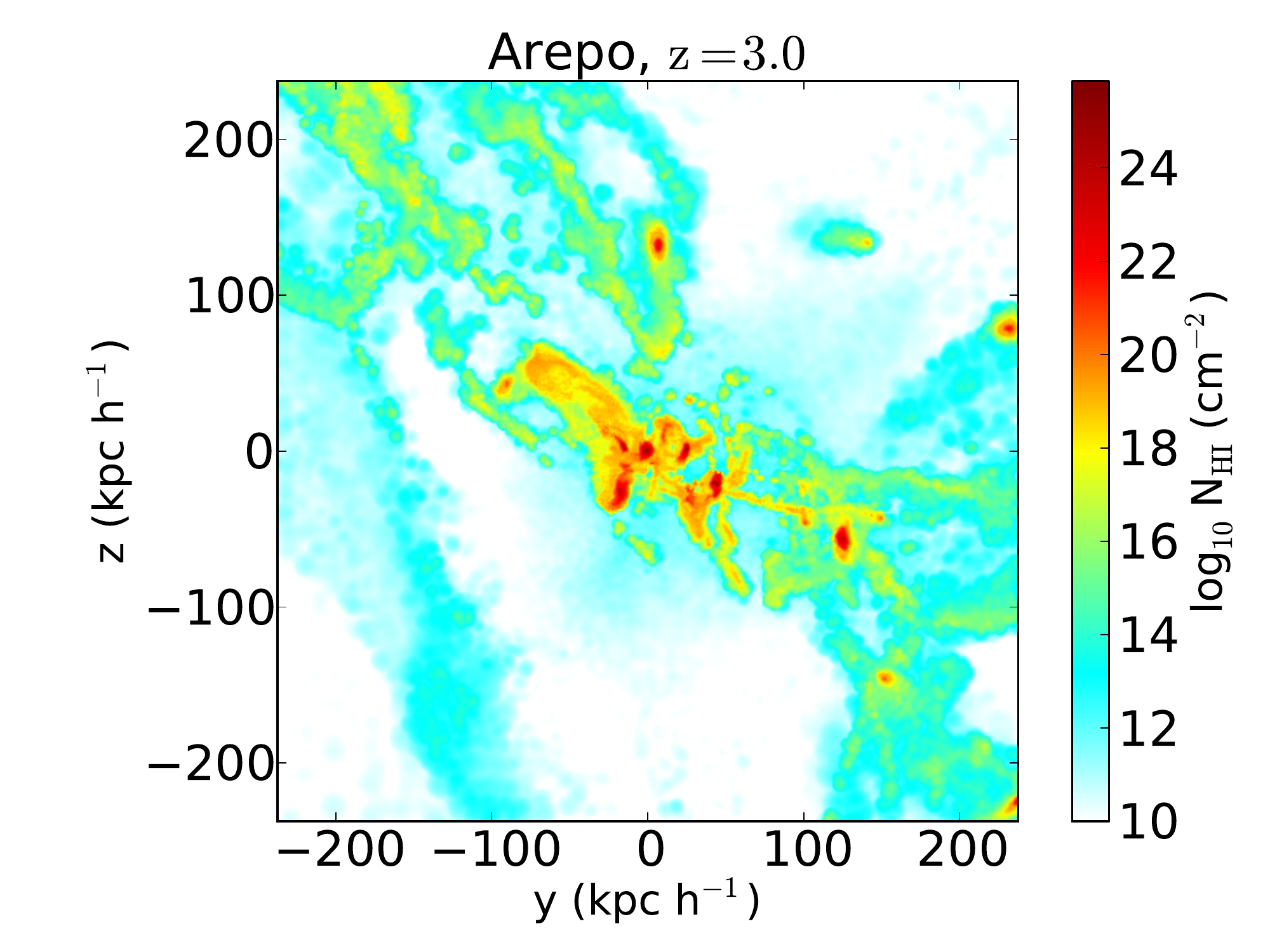}
\includegraphics[width=0.45\textwidth]{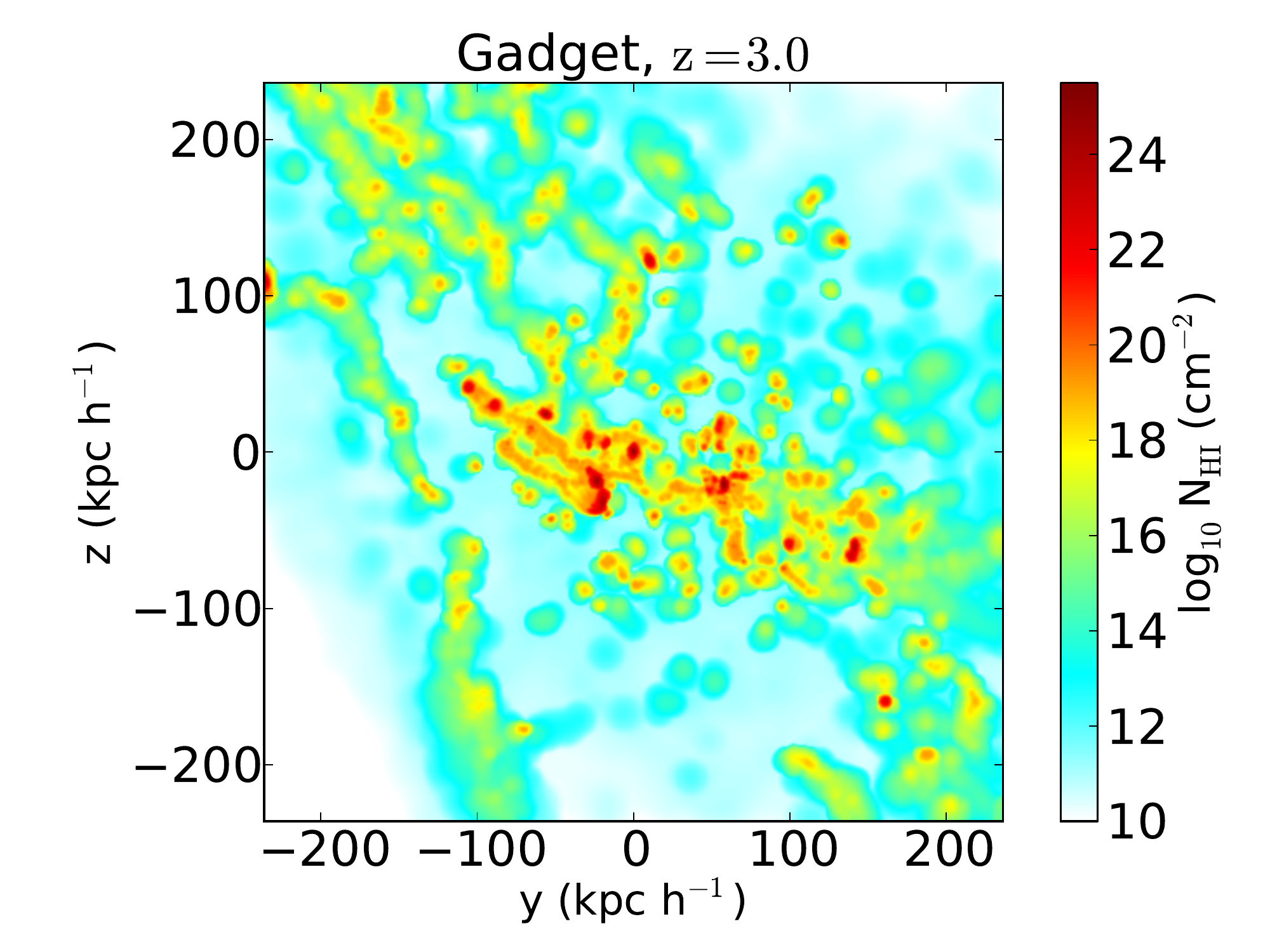} \\
\includegraphics[width=0.45\textwidth]{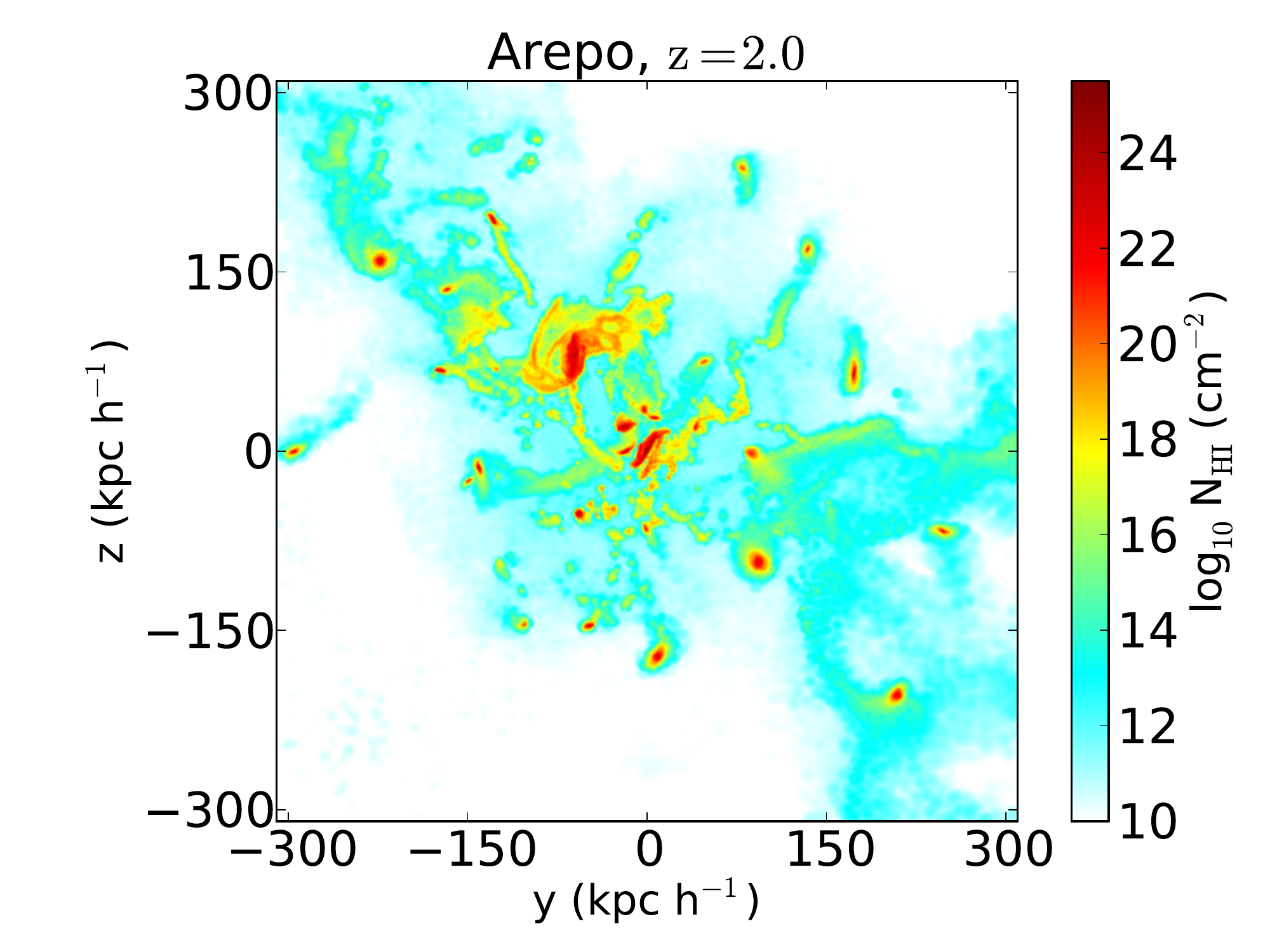}
\includegraphics[width=0.45\textwidth]{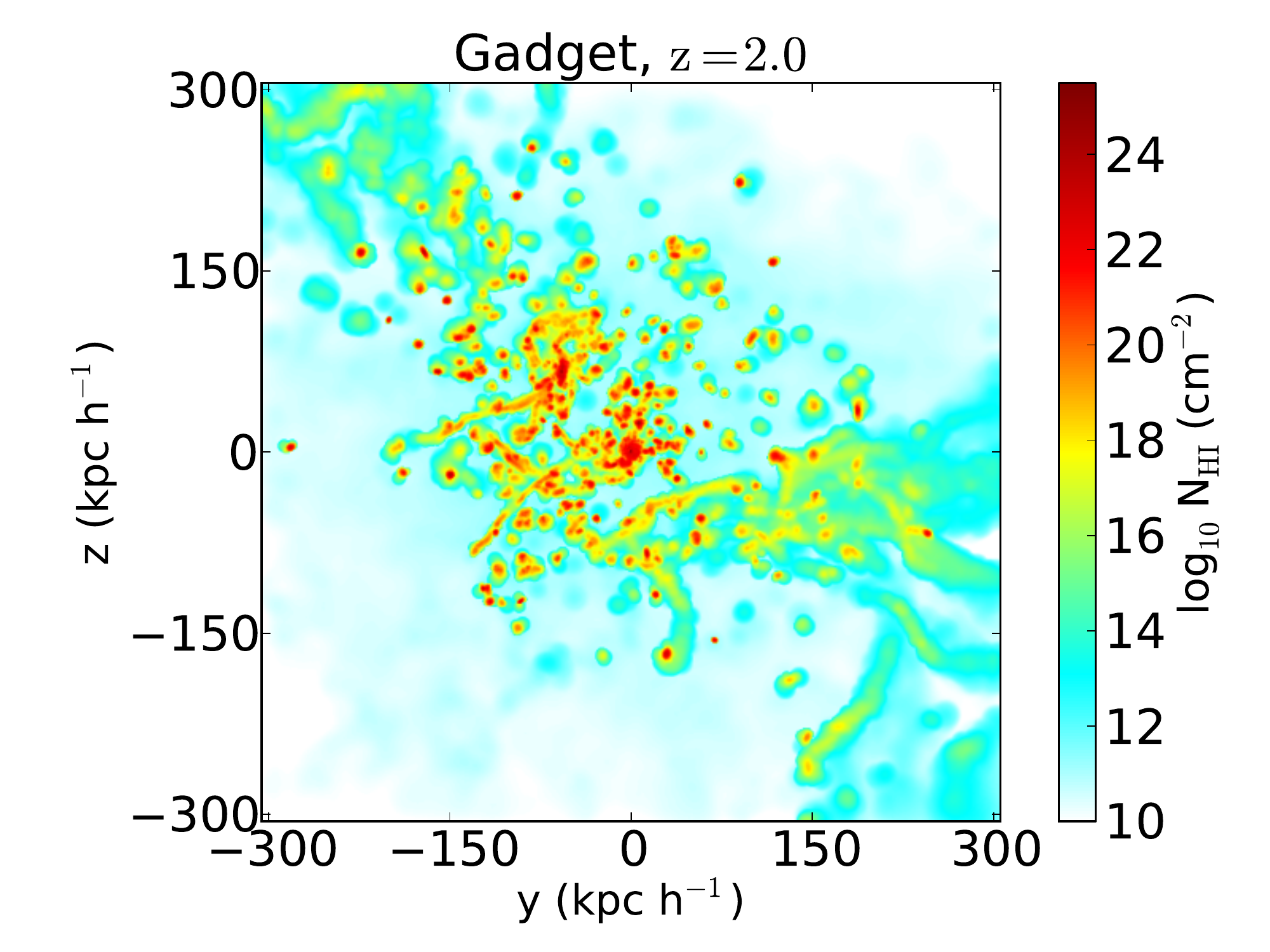}
\caption{Neutral hydrogen distribution around the largest halo, at three different redshifts.
Each row shows a different redshift, from top to bottom, $z=4$, $3$, and $2$. 
The left column shows \arepo\ and the right \gadget. $y$ and $z$ are comoving coordinates. 
}
\label{fig:prettyhalo}
\end{figure*}

\begin{figure*}
% \centering
\includegraphics[width=0.45\textwidth]{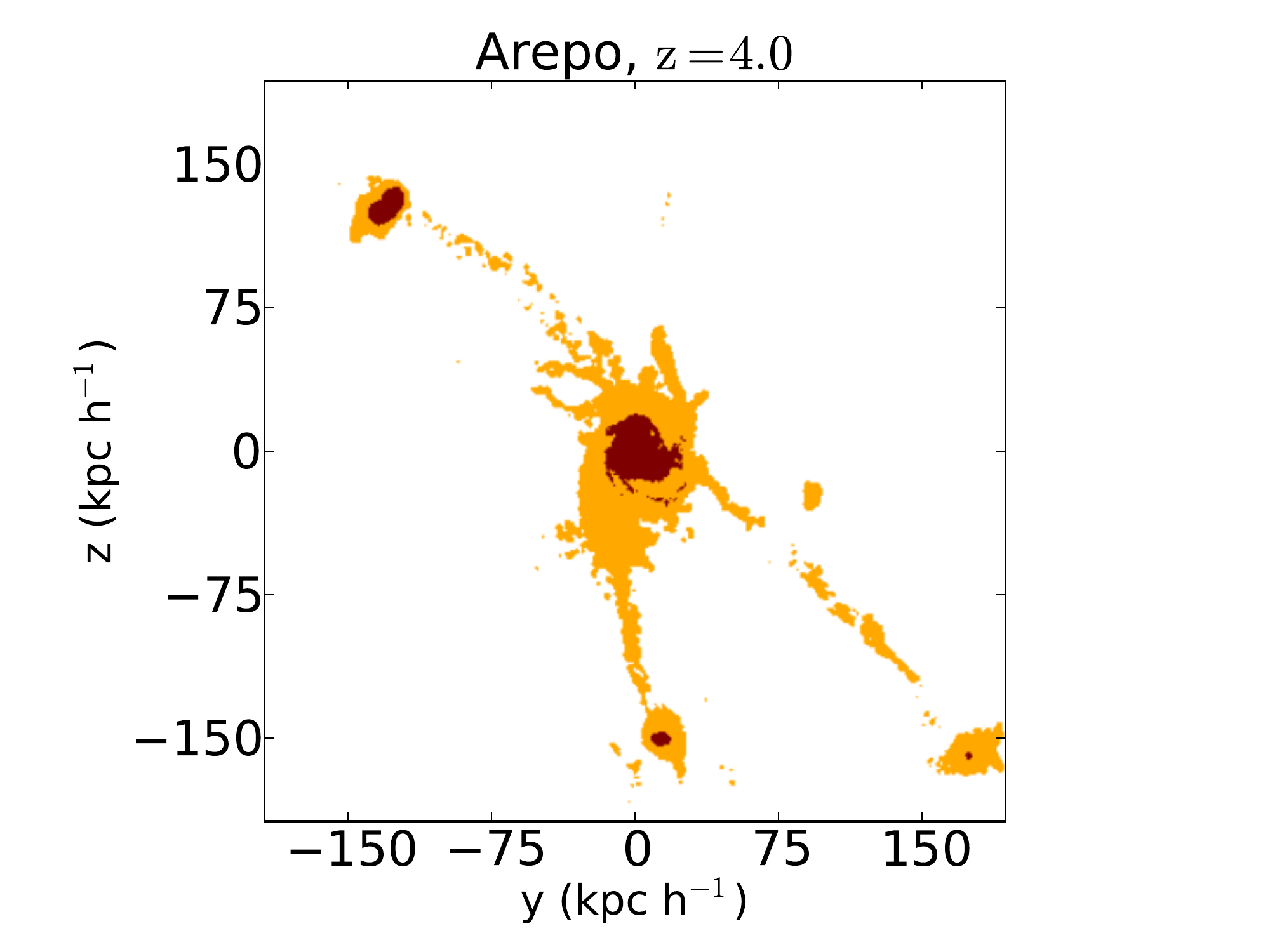}
\includegraphics[width=0.45\textwidth]{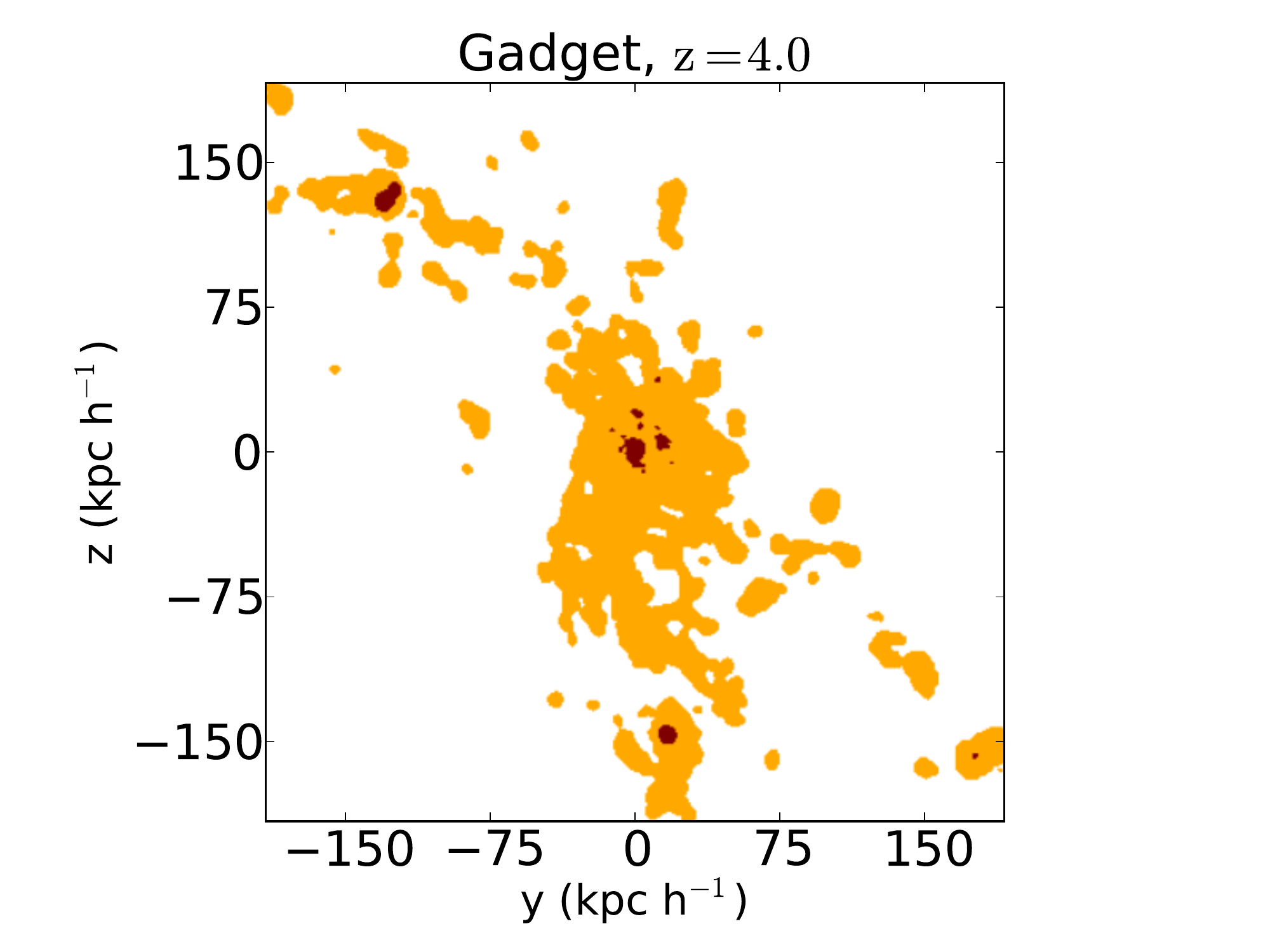} \\
\caption{Distribution of gas around the most massive halo at $z=4$, highlighted so that all DLA cells 
are in red and all LLS cells are in orange. Lower density cells are in white.
The left panel shows \arepo, the right panel \gadget. $y$ and $z$ are again comoving.
}
\label{fig:prettycuthalo}
\end{figure*}

Figure \ref{fig:prettyhalo} shows the distribution of neutral hydrogen in
the largest halo in our simulation for $z=4-2$. 
For column densities with $N_\mathrm{HI} > 10^{19}$ \NHunit, gas is almost entirely
self-shielded and thus is traced extremely well by the neutral hydrogen.
We can see that \gadget~produces a large number of small, circular, gaseous artefacts,
which are largely absent in \arepo. Similar ``blobs'' have been seen
in other cosmological SPH simulations (including calculations with
different SPH codes), and we interpret their existence as a numerical
artefact due to the suppression of fluid instabilities and mixing in
SPH; see \cite{Torrey:2012} for further details.
In \gadget, these SPH blobs make up the bulk of the LLS, especially 
in the column density range $10^{17} < N_\mathrm{HI} < 10^{19}$ \NHunit.
\arepo~reveals that LLS are more commonly produced from distinct filamentary structures.
Figure \ref{fig:prettycuthalo} shows this more explicitly. Here we have
discretised the column density map; any cell with $N_{\rm HI} <
10^{17}\,{\rm cm}^{-2}$ is shown in white, cells producing LLS are
orange, while DLAs are red.

There is a more subtle change in the DLA cross-section.  DLAs in
\arepo\ are more concentrated in the centre of the halo, but the
overall cross-section is not significantly changed; although there
are fewer substructures, this is partially compensated by the
higher accretion rate of the central halo.

\subsection{Small halos} 
\label{sec:centralden}

\begin{figure}
\includegraphics[width=0.45\textwidth]{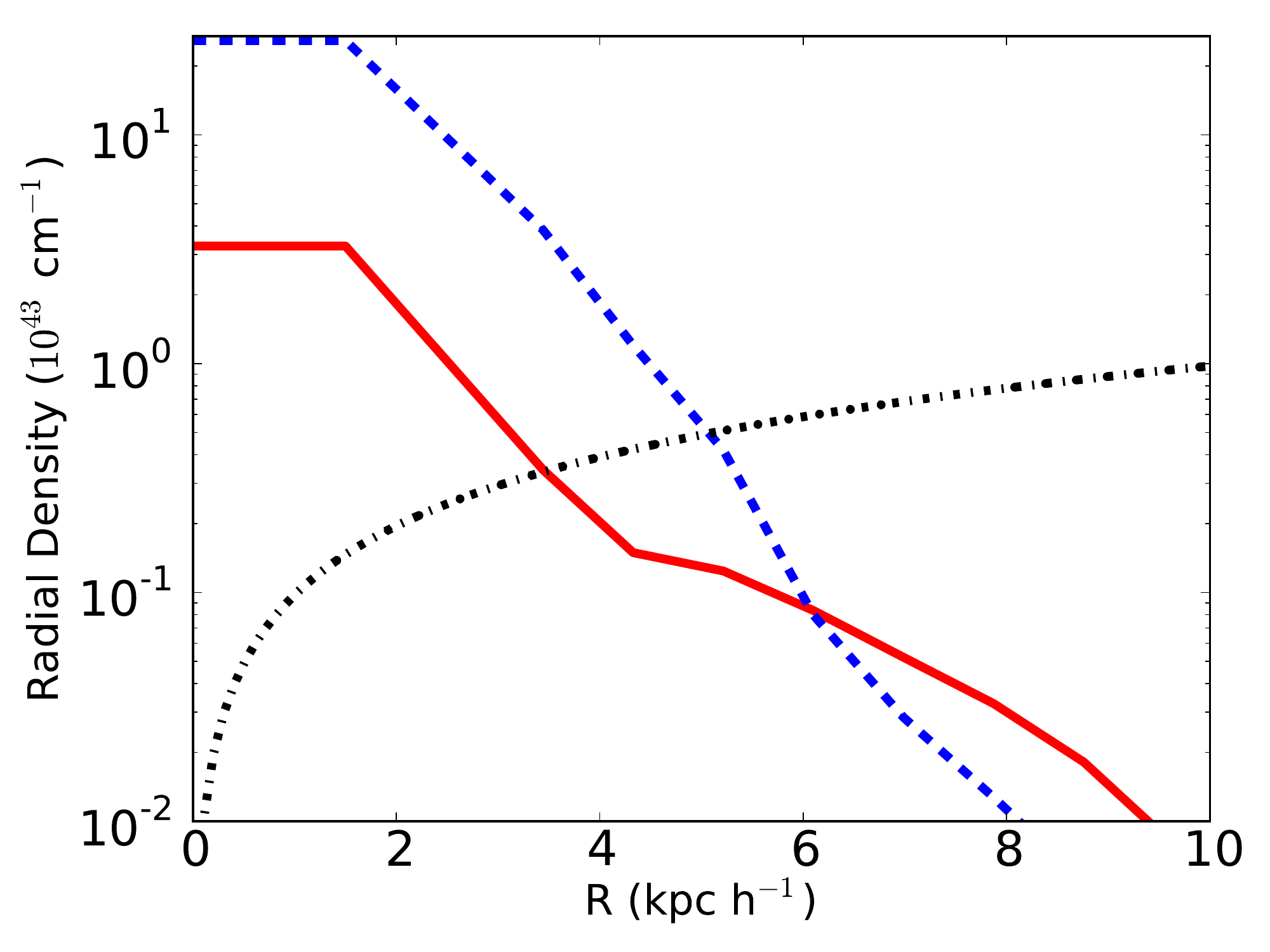}
\caption{Stacked radial density profiles around the centre of halos
  with mass $3 \times 10^9 \Msun < M < 3.5 \times 10^9 \Msun$ at
  redshift $z=3$.  The blue dashed lines show $662$ stacked halos from \gadget, while
  the red solid lines show $719$ stacked halos from \arepo. The black dot-dashed
  line indicates the density required for being over the DLA cut-off,
  $N_{\rm HI} = 10^{20.3}\,{\rm cm}^{-2}$. $R$ is comoving, but the 
  radial density is in physical units.
}
\label{fig:radialhalo}
\end{figure}

Figure \ref{fig:radialhalo} shows the radial profile obtained by stacking all halos with mass 
$3 \times 10^9 \Msun < M < 3.5 \times 10^9 \Msun$  at $z=3$. 
We radially integrated the neutral hydrogen grids shown in Figure~\ref{fig:prettyhalo}, 
and averaged over all halos in the given mass range. These halos are 
approximately spherical, so the effect of substructure is relatively small. 
The column density in the central $5 \kpch$ of the halo is larger by a factor of about $7$ in \gadget.
This effect is more pronounced at redshift $z=4$ (a factor of $10$) than at $z=2$ (a factor of $3$). 
\edit{This does not appear to be due to resolution effects}; the density of the innermost $5 \kpch$ is larger in \gadget\ 
for all halos with mass up to $M = 5 \times 10^{10} \Msun $, in both the $2
\times 512^3$ and $2 \times 256^3$ particle simulations.
However, for $M > 10^{10} \Msun$ (and at $z=2$), the characteristic size of a DLA is 
much larger than $5 \kpch$, so changes in the central density have a much smaller effect. 

\subsection{DLA and LLS cross-section}
\label{sec:sigma_dla}

%%%%%%%%%%%%%%%%%%%%%%%%%%%%%%%%%%%%%%%%
\begin{figure*}
% \centering
\subfigure{
\includegraphics[width=0.45\textwidth]{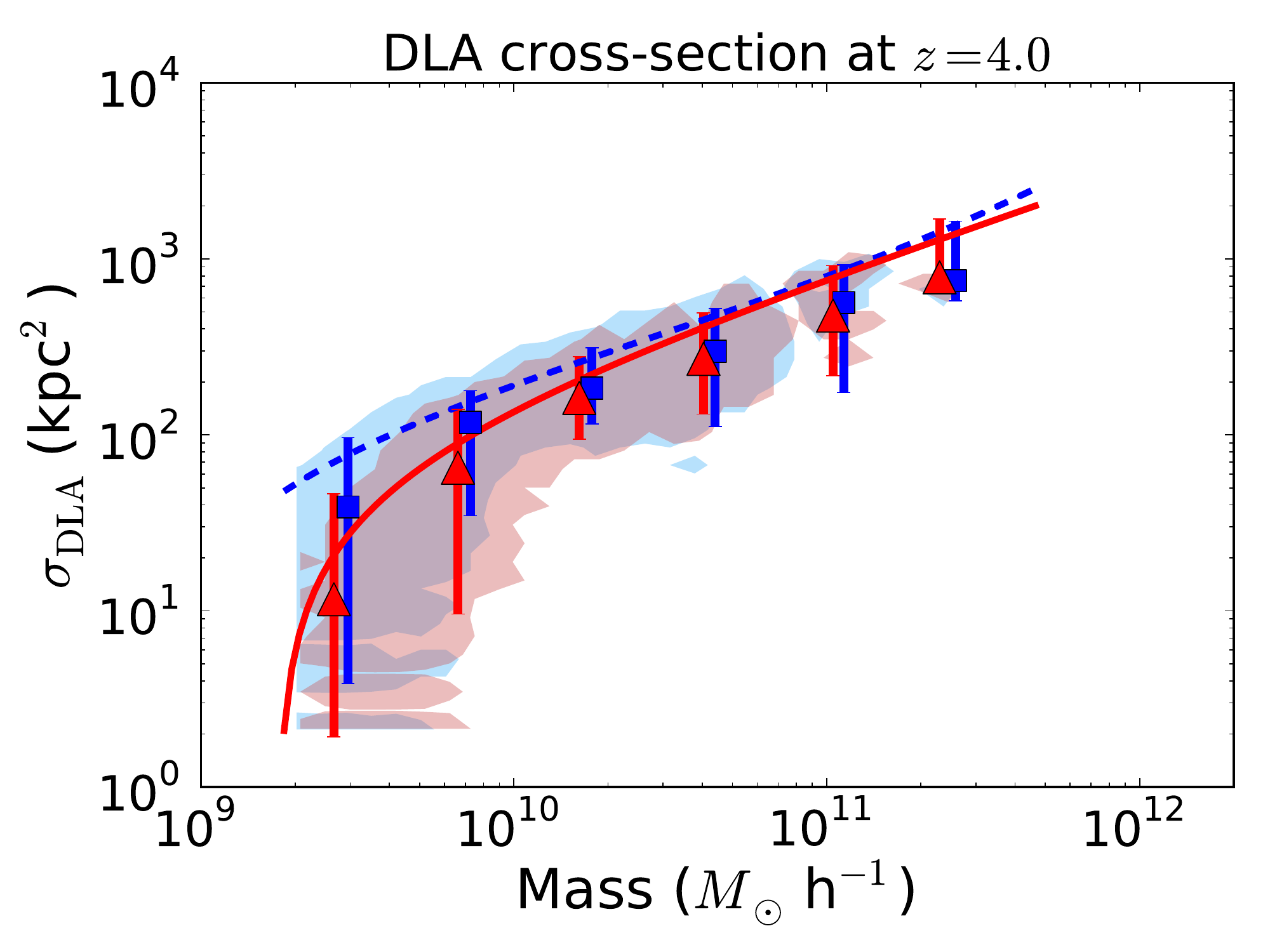}
\includegraphics[width=0.45\textwidth]{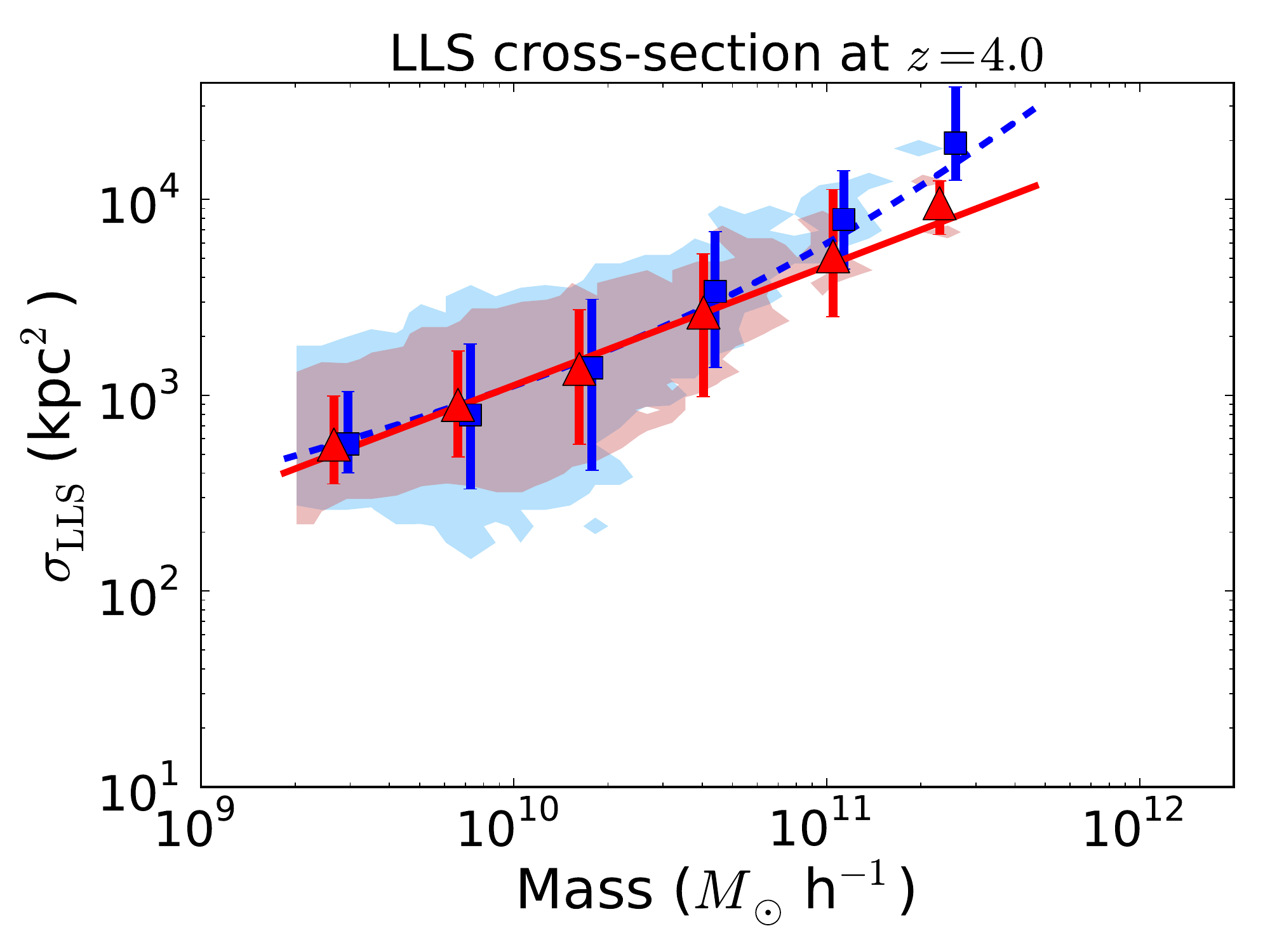}
\label{fig:sigmaDLA_z4}
} \\
\subfigure{
\includegraphics[width=0.45\textwidth]{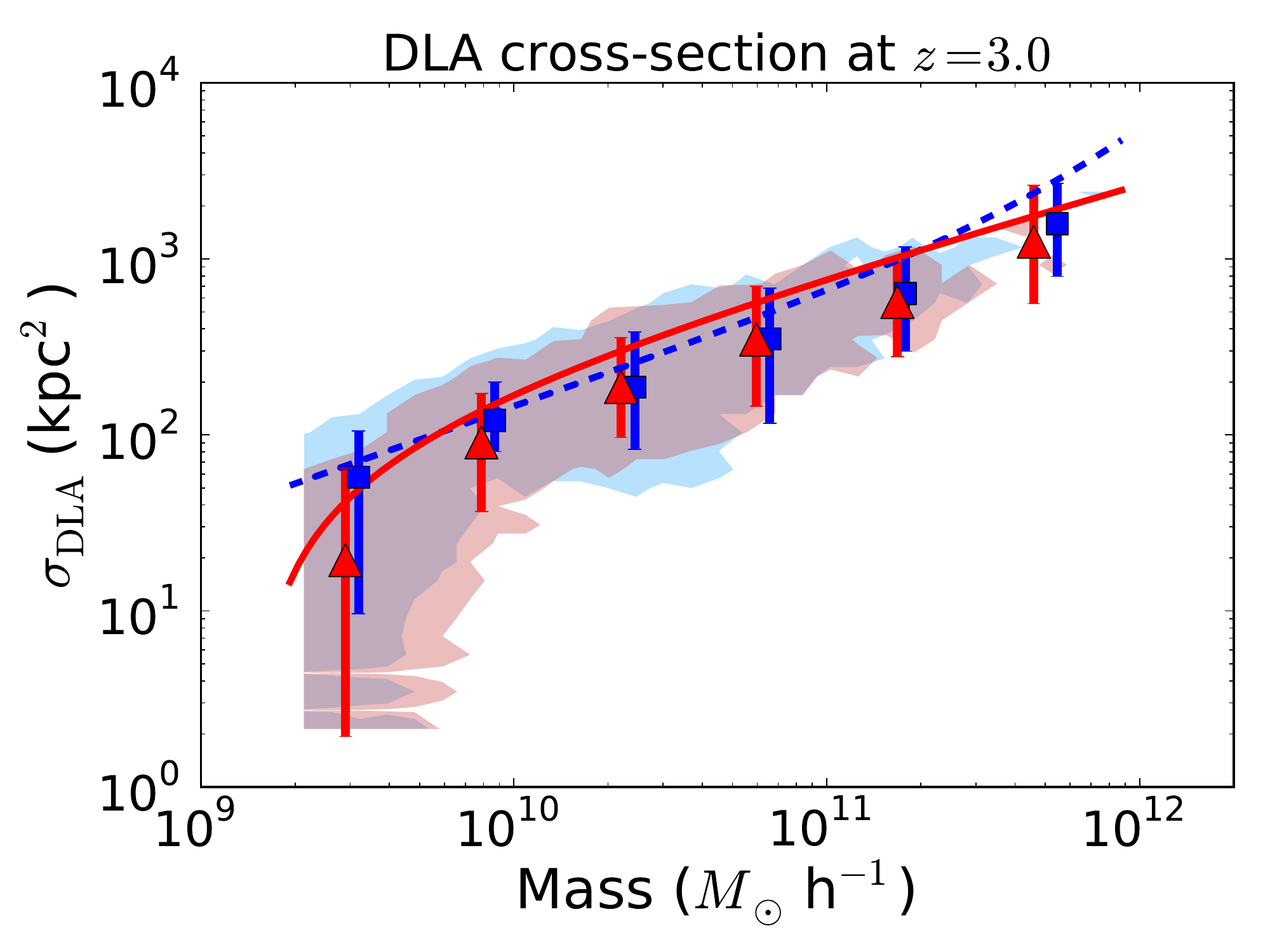} 
\includegraphics[width=0.45\textwidth]{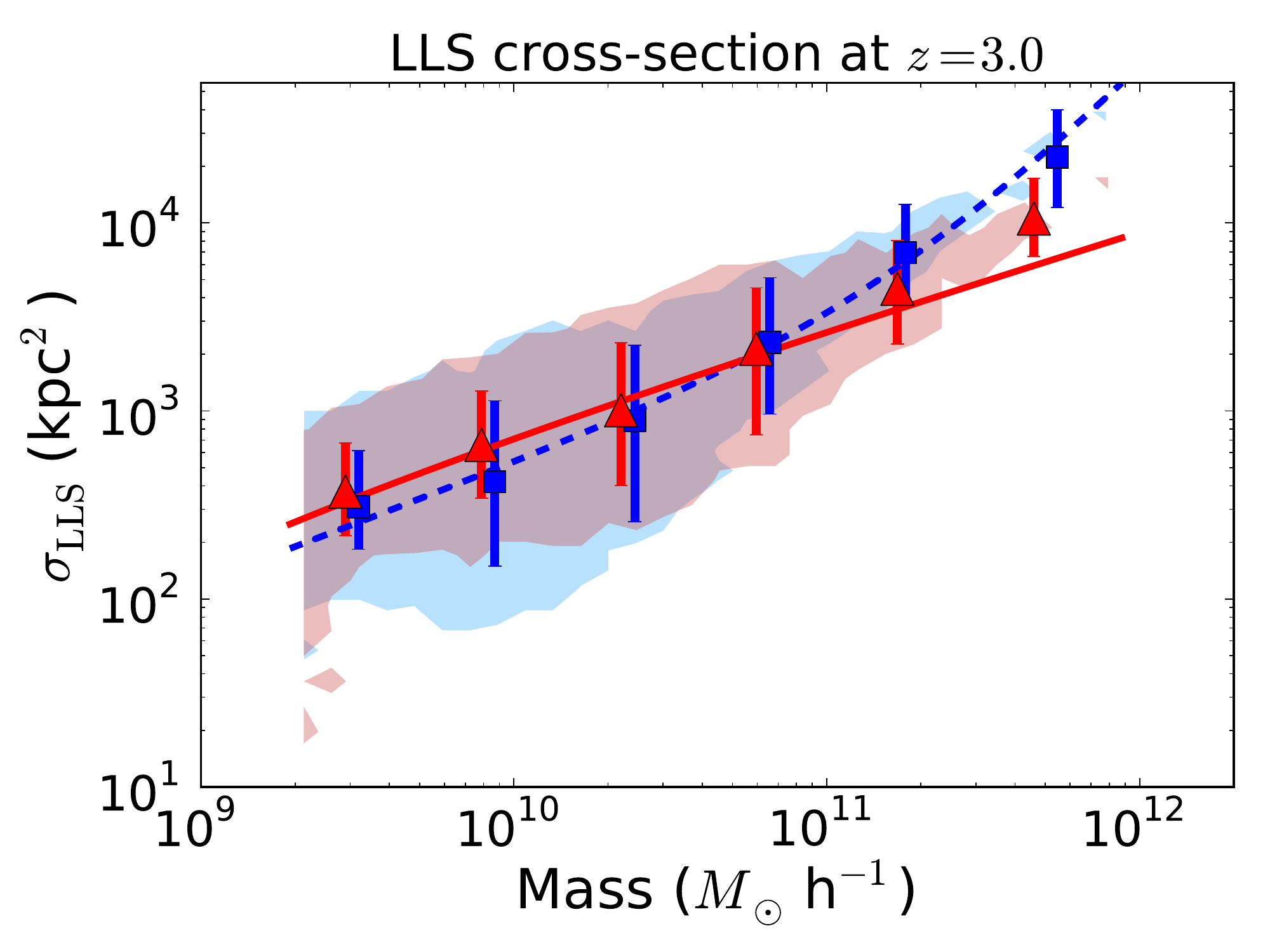} 
\label{fig:sigmaDLA_z3}
} \\
\subfigure{
\includegraphics[width=0.45\textwidth]{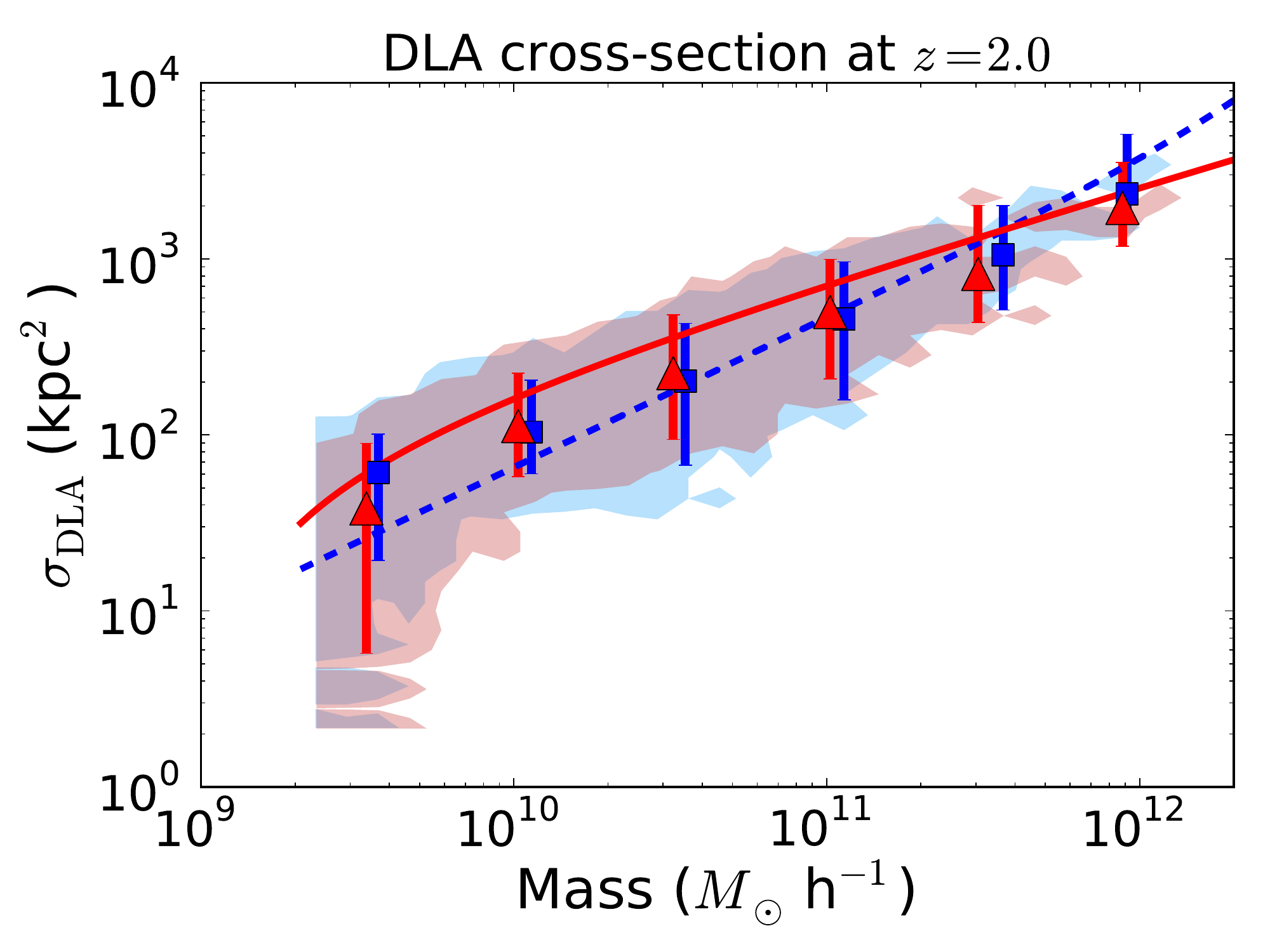}
\includegraphics[width=0.45\textwidth]{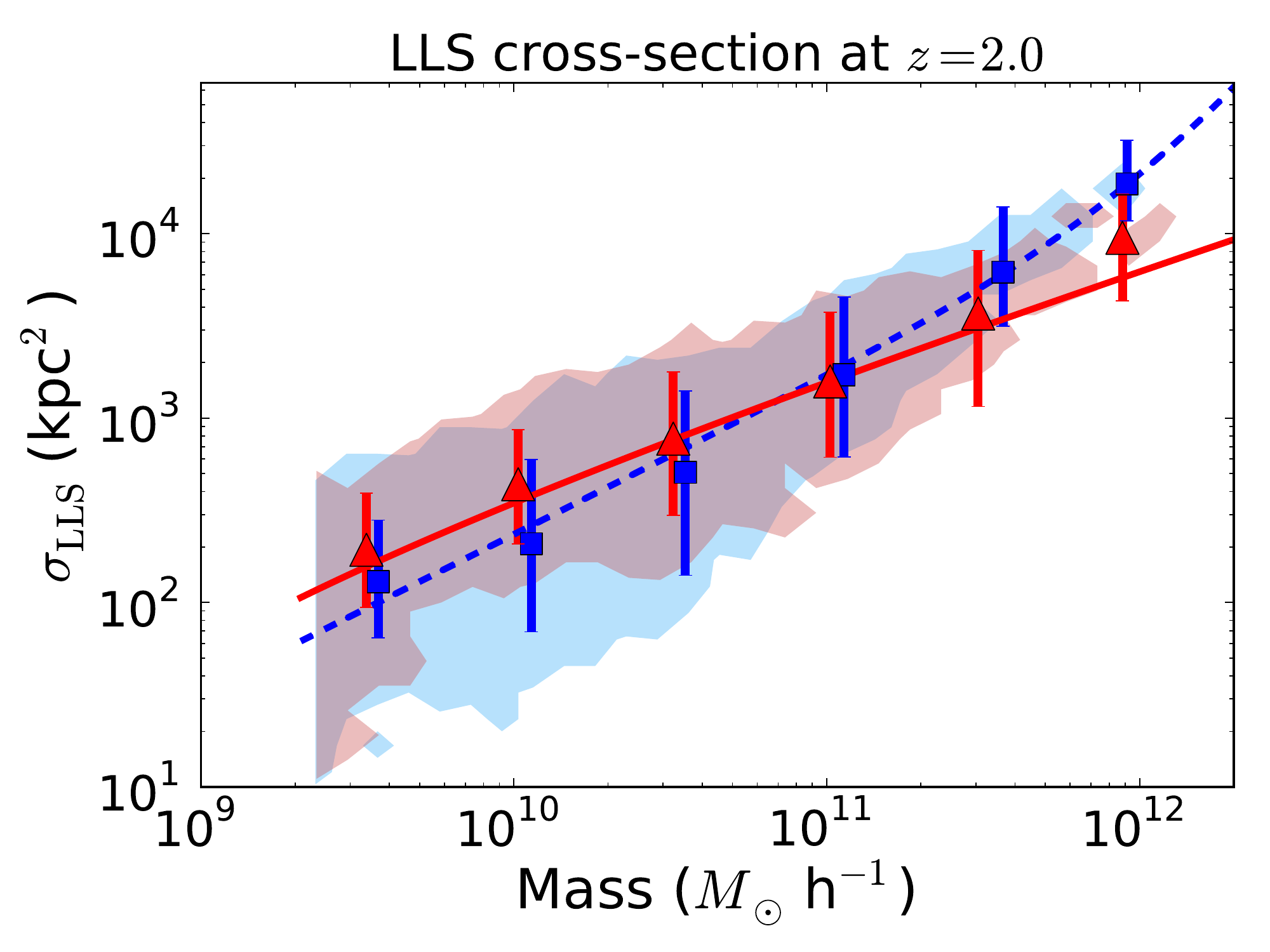}
\label{fig:sigmaDLA_z2}
} \\
\caption{{\em Left panels:} The comoving DLA cross-section, $\sigmaDLA$. {\em
    Right panels:} The LLS cross-section, $\sLLS$. Each row shows a
  different redshift, $z=4$ (top), $z=3$ (middle), and $z=2$ (bottom).
  Regions containing at least one \gadget\ halo are shown in blue, while
  regions containing at least one \arepo\ halo are shown in red.  The
  symbols with error bars show the median DLA cross-section in seven
  evenly-spaced mass bins; the error bars show the upper and lower
  quartiles. Red triangles are for \arepo, while blue squares are for
  \gadget.  The symbols for \gadget~have been offset horizontally 
  by $10$\% for visibility.
  The lines denote the results of our fit; red solid lines are for
  \arepo, while blue dashed lines are for \gadget.  
}
\label{fig:sigma_DLA}
\end{figure*}

%TABLE
\begin{table}
\begin{center}
\begin{tabular}{|c|c|c|c|c|c|c|}
\hline
Redshift & Code & $a$ & $b$ & $c$ & $d$ & $e$ \\
\hline 
4   & Arepo   &  0.593  &  33.1  &  74.0  &  1500  &  1.07 \\ 
4   & Gadget   &  0.429  &  33.5  &  79.8  &  1070  &  2.97 \\ 
3   & Arepo   &  0.496  &  33.8  &  101  &  -474  &  0.529 \\ 
3   & Gadget   &  0.625  &  32.7  &  -27.3  &  316  &  2.73 \\ 
2   & Arepo   &  0.518  &  33.6  &  66.7  &  -1120  &  0.787 \\ 
2   & Gadget   &  0.849  &  31.5  &  -50.0  &  20.1  &  2.51 \\ 
\hline
\end{tabular}
\end{center} 
\caption{Numerical parameters of the fit described in Eq.~(\ref{eq:halo_fit}).
}
\label{tab:halofits}
\end{table}

%TABLE
We define the DLA cross-section, $\sigmaDLA$, of a halo to be the area covered by all grid cells with column density
$N_\mathrm{HI} > 10^{20.3}$ \NHunit. The LLS cross-section, $\sLLS$, is defined similarly, but 
with $10^{20.3} > N_\mathrm{HI} > 10^{17}$ \NHunit.

Figure \ref{fig:sigma_DLA} shows the DLA cross-section, $\sigmaDLA$,
as a function of halo mass. The shaded region delineates the area 
in the $M -\sigmaDLA$ plane populated by halos. Symbols with error bars
show the median, upper and lower quartiles of $\sigmaDLA$ for halos in seven
halo mass bins. In order to fit the features at small and large halo
masses, we modelled $\sigmaDLA$ and $\sLLS$ with a three-component power law
\begin{equation}
 \sigma = \left[\left(\frac{M}{M_0}\right)^a + \frac{d}{10^{N/5}} \left(\frac{M}{M_0}\right)^e\right] 10^{(b-N)/5} - c\,,
\label{eq:halo_fit}
\end{equation}
where $M_0 = 10^{10.5} \Msun$ and $N = 20.3$ for DLAs or $N=17$ for
LLS. The free parameters $a$, $b$, $c$, $d$, and $e$ are
found by a simultaneous fit to $\sigmaDLA$ and $\sLLS$\footnote{\edit{A consequence of fitting to both quantities 
together is that at $z=3$, the best-fit line matches $\sLLS$ but is slightly above the median $\sigmaDLA$.}}, 
using {\small MPFIT} \citep{Markwardt:2009}\footnote{As ported
to python by Mark Rivers and Sergey Koposov}. The resulting numerical
values are listed in Table \ref{tab:halofits} and the fit shown by 
the lines in Figure \ref{fig:sigma_DLA}. This procedure is similar
to that outlined in \cite{Nagamine:2004a}, except that we have
excluded halos which do not form DLAs from our fit, instead of
assigning them an arbitrary small cross-section as in that work.

Qualitatively, our results for $\sigmaDLA$ in \gadget~are in good
agreement with those of \cite{Nagamine:2004a} for overlapping halo
masses. They found that, for SPH, a turnover in the
$\sigmaDLA$-halo mass relation occurred for $M \sim 10^{8.5} \Msun$,
below the resolution limit of our simulations. They also found a slight
excess in $\sigmaDLA$ over their bare power law for large halos, which
we have fit for explicitly. Our $\sigmaDLA$ has a somewhat smaller
amplitude than the O3 model of \cite{Nagamine:2004a} and 
the $\sigmaDLA$-halo mass relation is somewhat shallower in slope.
This is in agreement with the results of \cite{Tescari:2009} and 
is probably due to a difference in cosmological parameters; 
\citet{Nagamine:2004a} used $\sigma_8 = 0.9$, while we have $\sigma_8 = 0.8$.

Halos in the mass range $10^{11}\Msun > M > 10^{10}\Msun$ show
similar cross-sections in both \arepo~and \gadget.
The extra substructure discussed in Section \ref{sec:haloshape} 
leads to much larger $\sLLS$ in \gadget~for halos of mass larger
than $10^{11} \Msun$, especially at low redshift, because of the 
increased number of massive halos. This effect is
smaller for $\sigmaDLA$, as discussed above, but still present. 

Due to the effect discussed in Section \ref{sec:centralden}, 
$\sigmaDLA$ is significantly reduced in \arepo~for small halos, producing a
turnover in the power law relation between mass and DLA
cross-section. This turnover occurs because halos below a certain size
have a central column density which is on average less than the DLA
cut-off. It also occurs in \gadget, but at halo masses below the
resolution limit of our simulations \citep{Tescari:2009}. Notice that
this reduction in the cross-section of low-mass DLAs is not carried
over to the cross-section of low-mass LLS; if anything, small halos in
\gadget~have slightly lower $\sLLS$ than their counterparts in
\arepo. This suggests that these small halos have approximately the
same amount of neutral hydrogen in both codes, but that this gas is
more concentrated in the central peak in \gadget.

While the overall amplitude of $\sigmaDLA$ remains roughly constant
with redshift, that of $\sLLS$ decreases fairly substantially with
time, in both codes. Figure \ref{fig:prettyhalo} shows the reason for this;
by $z=2$, most of the filaments and streams that 
dominate the LLS cross-section have been swept into halos.

\subsection{Incidence rate of DLA systems}
\label{sec:dla_abund}

%%%%%%%%%%%%%%%%%%%%%%%%%%%%%%%%%%%%%%%%%
\begin{figure}
\includegraphics[width=0.45\textwidth]{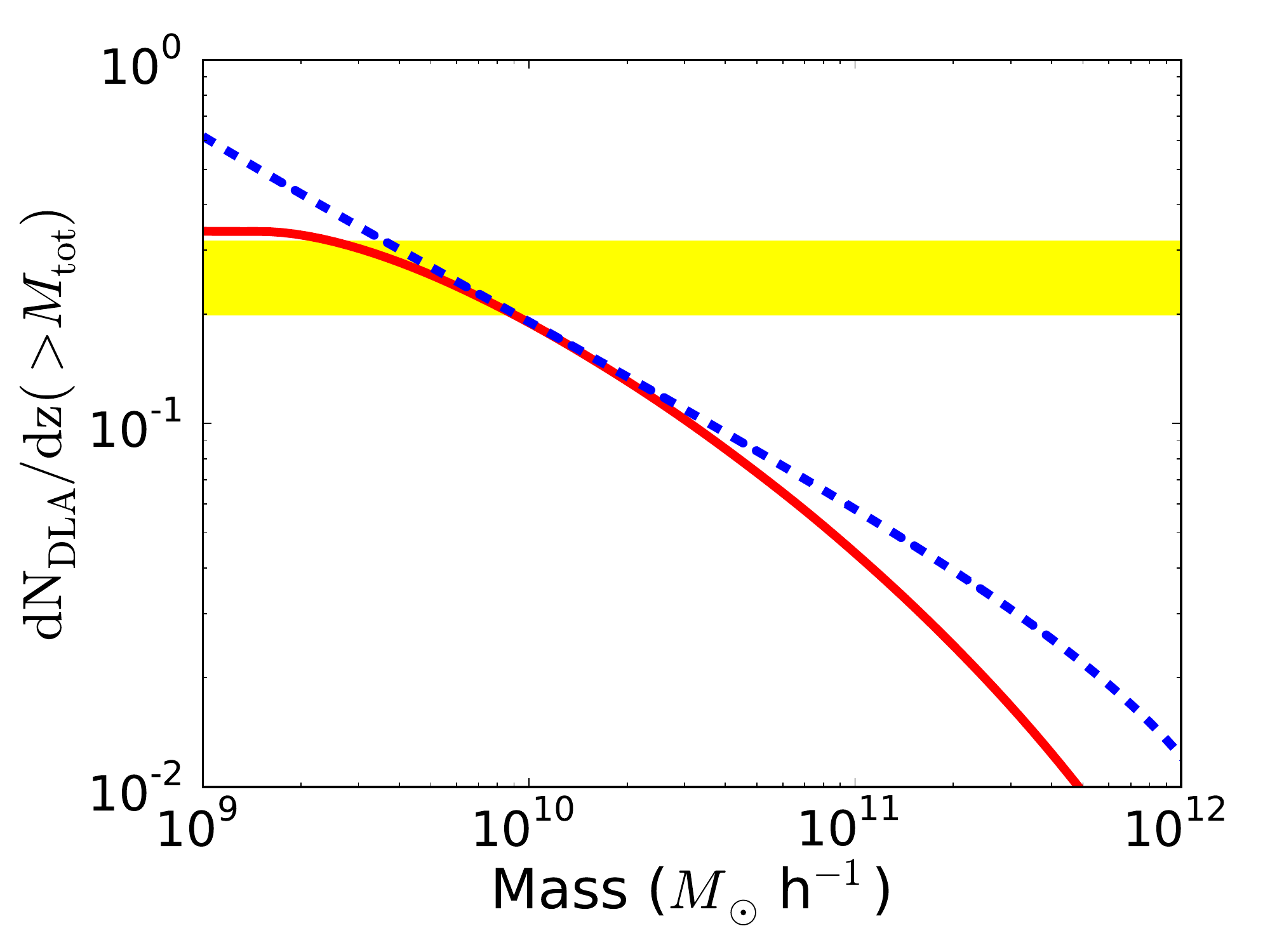}
\caption{Cumulative abundance of DLAs per unit redshift as a function of total halo mass at $z=3$ for \arepo (red) and \gadget (blue). 
The yellow bar shows the observed total cumulative DLA abundance of \protect\cite{Prochaska:2005}. 
}
\label{fig:dndZ}
\end{figure}
%%%%%%%%%%%%%%%%%%%%%%%%%%%%%%%%%%%%%%%%%

Following \cite{Nagamine:2004a}, we calculate the incidence of DLAs by convolving the halo mass 
function with our fit to $\sigmaDLA$. This allows us to account for 
dark matter halos of smaller mass than the resolution limit of the simulation. 
The cumulative number of DLAs per unit redshift is then defined as
\begin{equation}
  \frac{{\rm d} N_\mathrm{DLA}}{{\rm d}z} ( > M, z) = \frac{{\rm
      d}r}{{\rm d}z} \int_M^\infty \frac{{\rm d} n_h}{{\rm d} M} \sigma_{\rm DLA} {\rm
    d} M\,,
\end{equation}
where ${{\rm d} n_h}/{{\rm d} M}$ is the Sheth-Tormen dark matter halo
mass function, $\sigma_{\rm DLA}$ is given by Eq. (\ref{eq:halo_fit}),
truncated at $M = 10^{12.5} \Msun$ and
\begin{equation}
\frac{{\rm d} r}{{\rm d}z} = \frac{c}{H_0} \sqrt{\Omega_m (1+z)^3 + \Omega_\Lambda}\,
\end{equation}
defines the line element.

Figure \ref{fig:dndZ} shows ${{\rm d} N_\mathrm{DLA}}/{{\rm d}z}$ for the \arepo\ and \gadget\ simulations.
The yellow band marks an observational estimate of the total DLA abundance, 
\begin{equation}
 \log_{10} \left(\frac{d N_\mathrm{DLA}}{dz}\right) = -0.6 \pm 0.1\,, 
\end{equation}
recovered by \cite{Nagamine:2007} from the data of \cite{Prochaska:2005}. 
Our results for the \gadget~simulation are similar to the weak wind scenario of \cite{Tescari:2009} and 
produce more DLAs than are observed. They showed that this discrepancy could be reduced with 
the addition of feedback processes. We find that the lower DLA cross-section of small and large 
mass halos in \arepo~combine to reduce the cumulative DLA abundance by a factor of two, bringing it 
almost into agreement with the upper limit from observations.

\subsection{Column density function}
\label{sec:columnden}

%%%%%%%%%%%%%%%%%%%%%%%%%%%%%%%%%%%%%%%%
\begin{figure*}
\includegraphics[width=0.33\textwidth,clip]{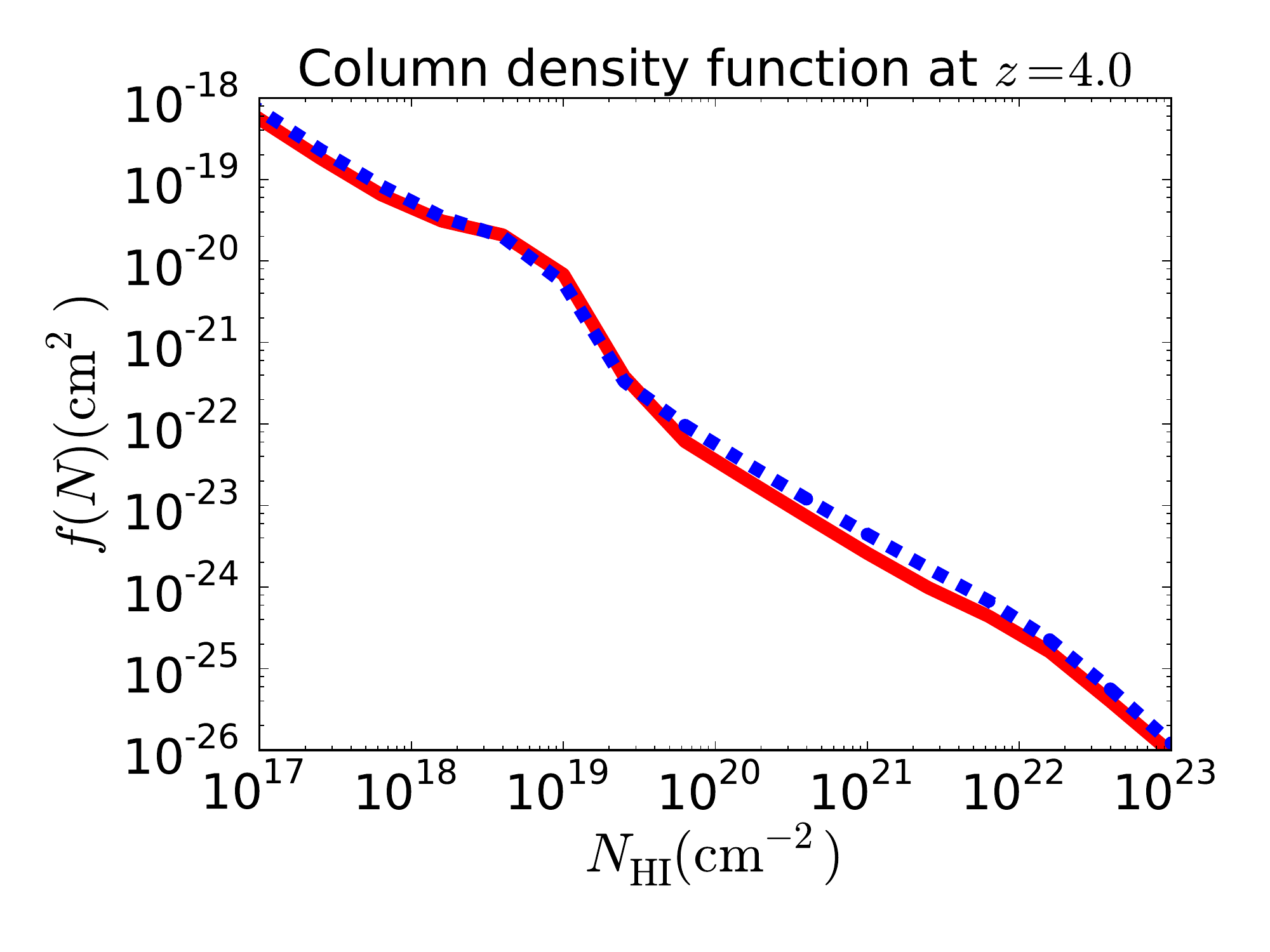}
\includegraphics[width=0.33\textwidth,clip]{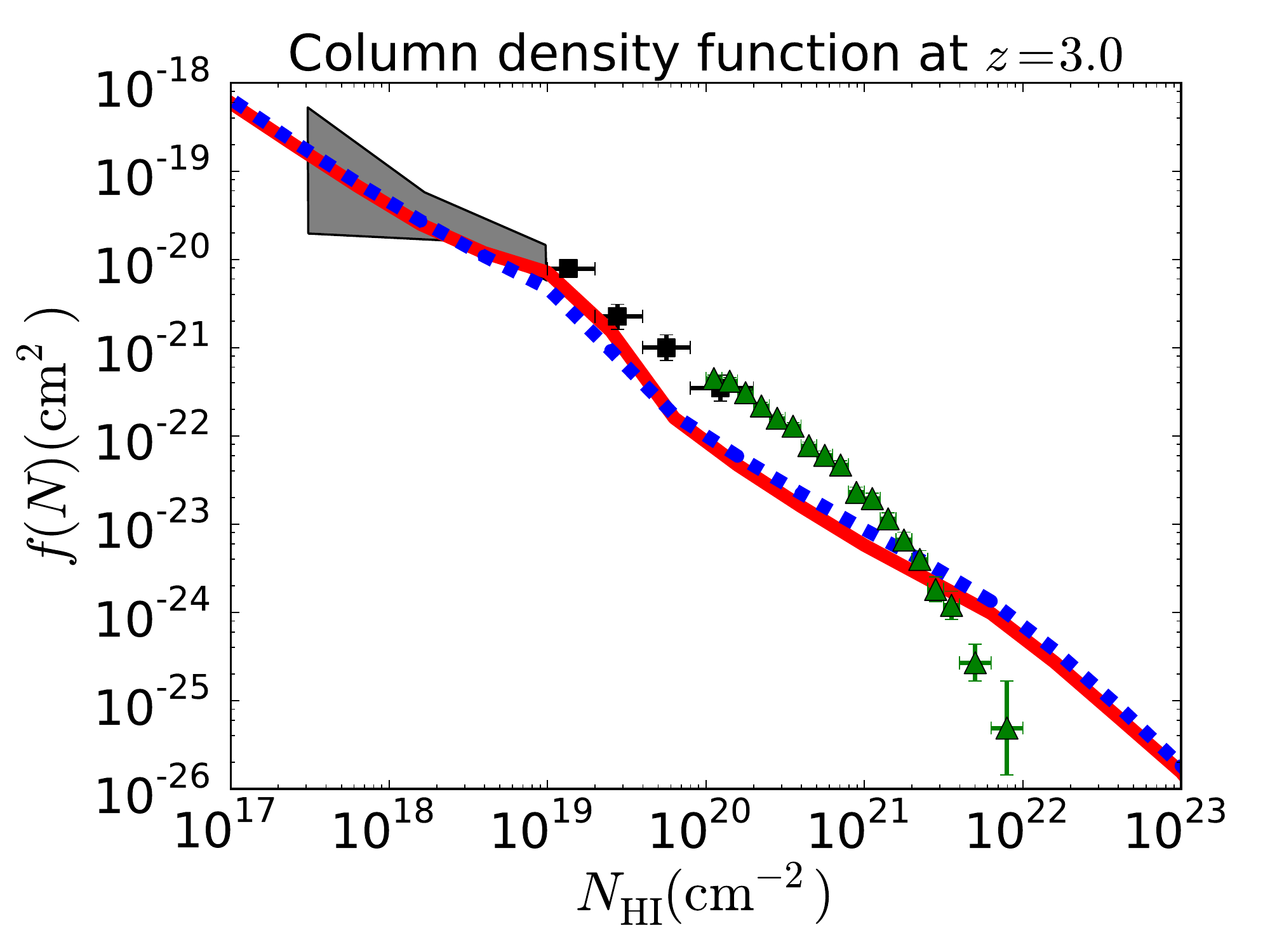}
\includegraphics[width=0.33\textwidth,clip]{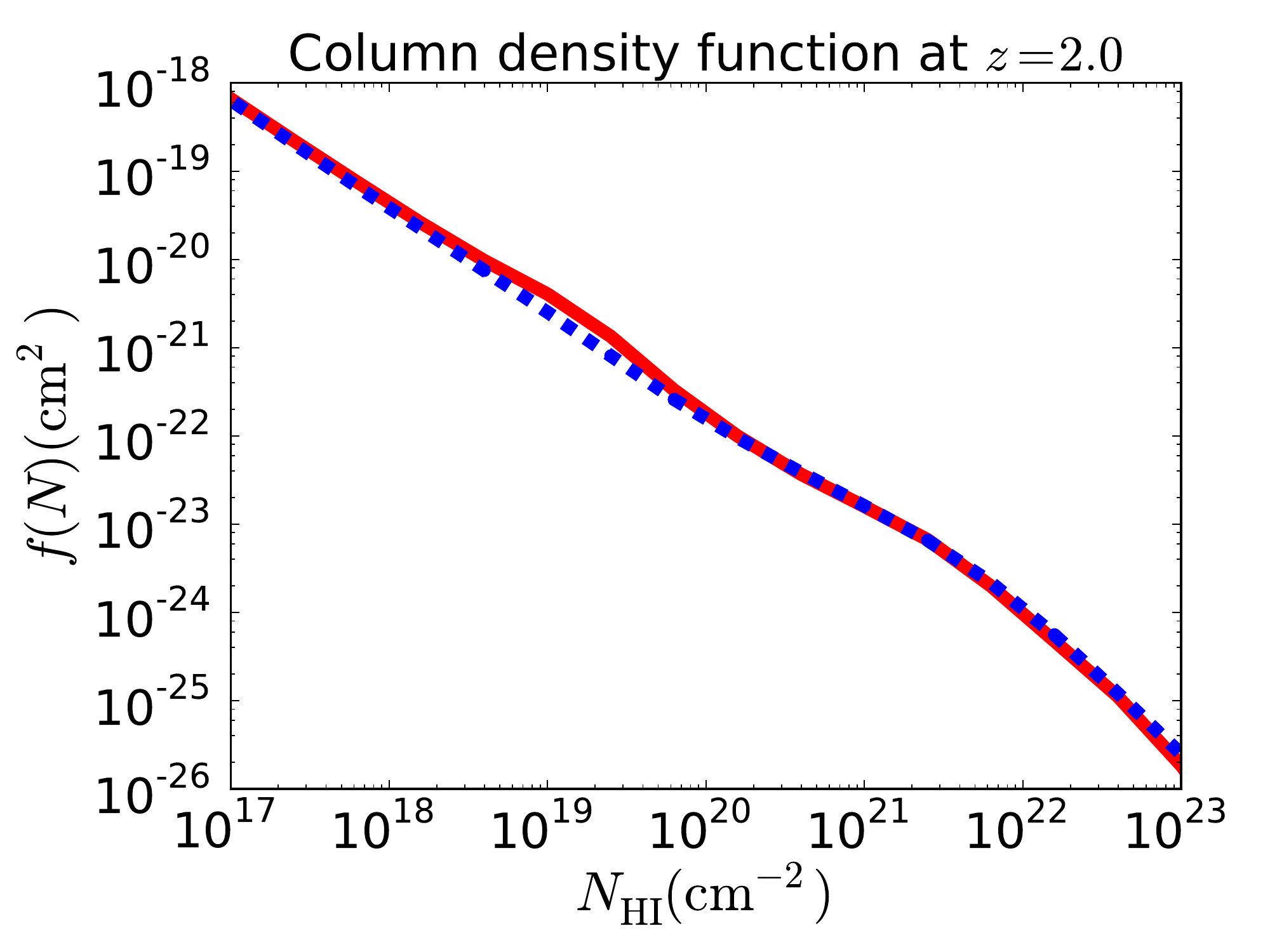} 
\caption{The column density function extracted from \gadget\
  simulations (blue dashed lines) and \arepo\ simulations (red solid
  lines).  Green triangles show the constraints of
\protect\cite{Noterdaeme:2009} at $z\sim 3$, 
black squares those of \protect\cite{OMeara:2007} at $z \sim 3.1$, and the grey region those of 
\protect\cite{Prochaska:2010} at $z=3.7$.
}
\label{fig:column}
\end{figure*}

Figure \ref{fig:column} shows the neutral hydrogen column density function, defined in Section \ref{sec:dlaobs}.
Our results suggest a significantly shallower column density function than preferred by observations. 
This was also seen by \cite{Nagamine:2004a} and \cite{Tescari:2009}, 
who found that adding feedback to the simulations produced better agreement. 
We will address this question in future
work based on forthcoming \arepo\ simulations with stronger wind feedback.

\begin{figure}
\includegraphics[width=0.45\textwidth]{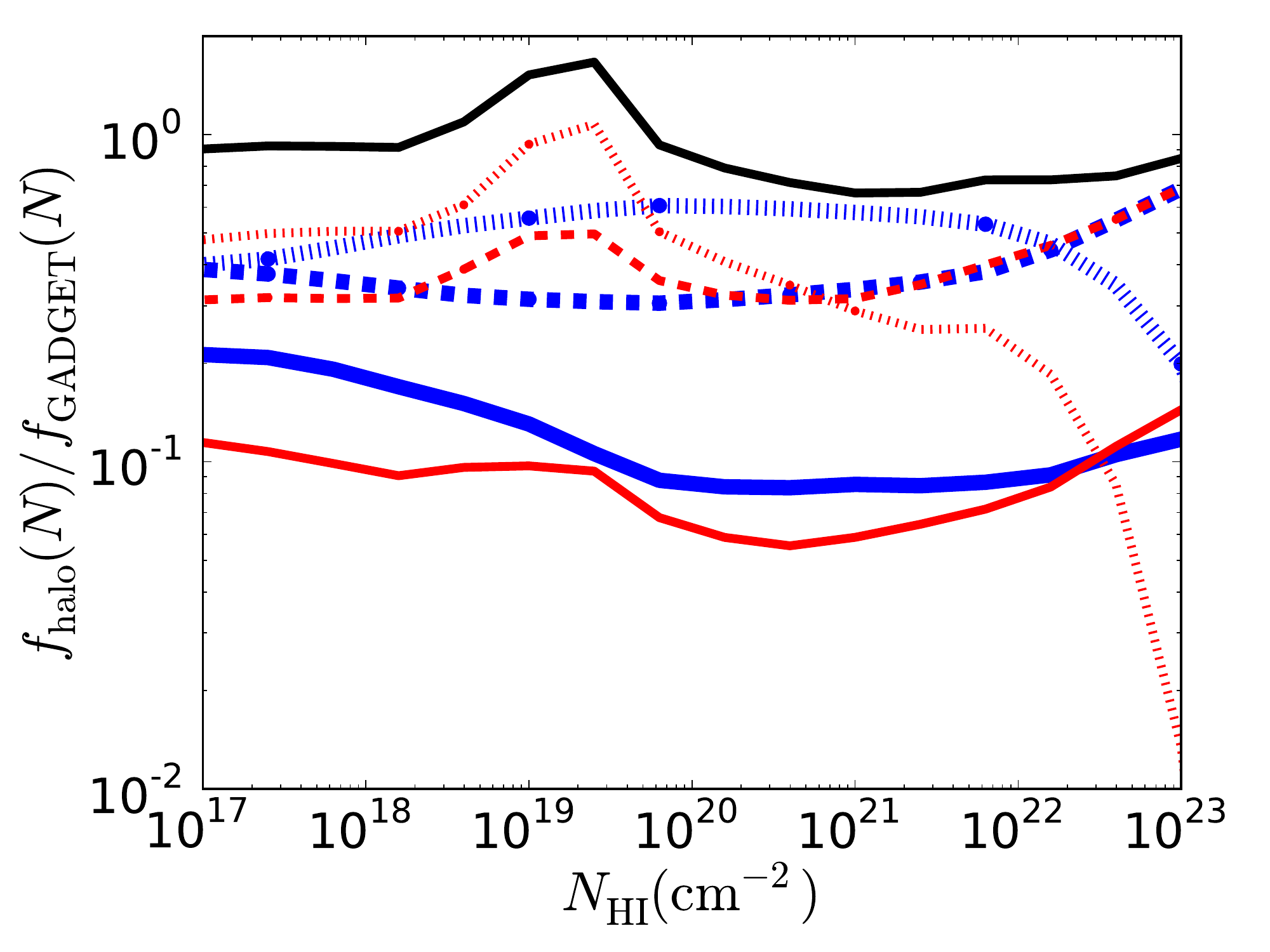}
\caption{The contribution to the column density function of halos with
  mass $M > 10^{11} \Msun$ (solid), $10^{11} \Msun > M > 10^{10}
  \Msun$ (dashed) and $M < 10^{10} \Msun$ (dotted), normalised by the
  total $f(N)$ for \gadget~at $z=3$.  Blue (thick) lines denote \gadget~and
  red (thin) lines \arepo. The uppermost solid black line shows the total column density function
  for \arepo divided by that for \gadget.
}
\label{fig:columnbreak}
\end{figure}

Overall, the differences in the column density function between the
codes are fairly limited; a change of $\sim 30\%$ of the total.  At first glance this
may seem puzzling; why have the generally smaller \arepo\ halos made so
little difference to the column density function?
Figure~\ref{fig:columnbreak} shows the fraction of the column density
function contributed by halos in three mass ranges, corresponding to
those in Sections \ref{sec:haloshape} and \ref{sec:centralden}.  The
column density function has been calculated only for those halos with
the desired mass and then normalised using the total \gadget\ column
density function. This allows us to clearly see the differences
between the codes.  The reduced substructure in $ M > 10^{11} \Msun$
halos does lead to a reduced column density for these halos at
$N_\mathrm{HI} \sim 10^{20}$ \NHunit, but their rarity means they make
up only for a small fraction of the total column density function.
The largest change is produced by the reduction in cross-section of $
M < 10^{10} \Msun$ halos, although this too is diluted because 
$10^{10} \Msun < M < 10^{11} \Msun$ halos, which make up almost as great a
fraction of the total column density, are mostly unchanged.  This
explains why differences become larger with increasing redshift; we
found in Section~\ref{sec:centralden} that the effect on $ M < 10^{10}
\Msun$ halos became larger at early times.

Both codes produce a slight kink at $10^{19} - 10^{20}$ \NHunit. 
This feature is more prominent in \arepo, and is a consequence of the
transition between self-shielding and UVB ionization equilibrium, made more prominent by 
the smoother gas distribution in \arepo. 

%%%%%%%%%%%%%%%%%%%%%%%%%%%%%%%%%%%%%%%%

%%%%%%%%%%%%%%%%%%%%%%%%%%%%%%%%%%%%%%%%

\subsection[The Lyman-alpha forest]{The Lyman-$\alpha$ forest}
\label{sec:lya}

%%%%%%%%%%%%%%%%%%%%%%%%%%%%%%%%%%%%%%%
\begin{figure}
% \centering
\includegraphics[width=0.45\textwidth]{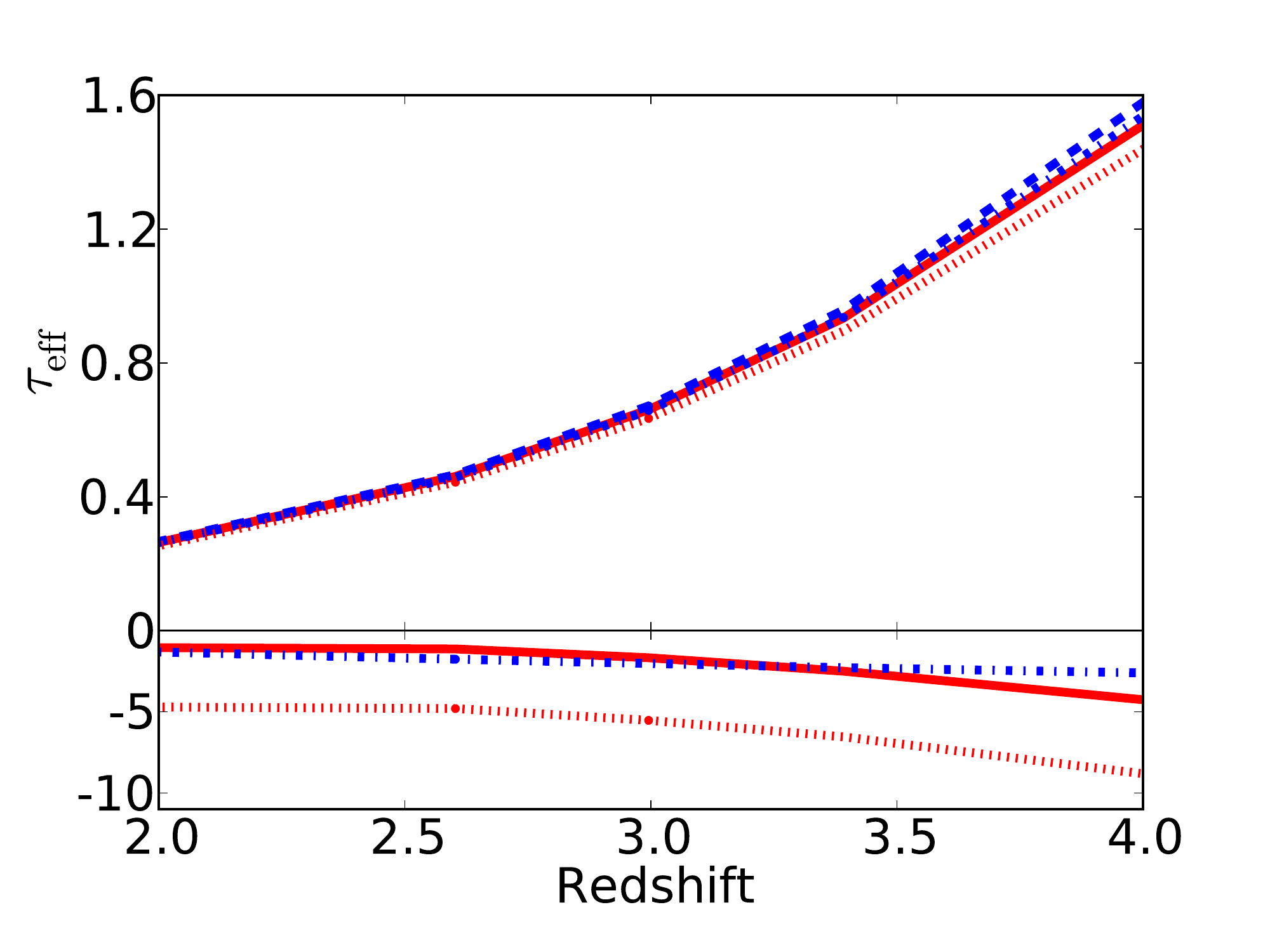}
\caption{The effective optical depth, $\tau_\mathrm{eff}$, as a
  function of redshift. Solid red lines show \arepo\ with $2\times
  512^3$ particles/cells, dashed blue lines show \gadget\ with $2\times
  512^3$ particles. Dot-dashed blue gives \gadget\ with $2\times 256^3$
  particles and dashed red shows \arepo\ with $2\times 256^3$
  resolution.  The bottom panel shows the percentage difference
  relative to the \gadget~run with $2\times 512^3$ particles.  }
\label{fig:tau_eff}
\end{figure}
%%%%%%%%%%%%%%%%%%%%%%%%%%%%%%%%%%%%%%%

%%%%%%%%%%%%%%%%%%%%%%%%%%%%%%%%%%%%%%%%
% \begin{figure}
% % \centering
% \includegraphics[width=0.45\textwidth]{plots/temp.pdf}
% \caption{The temperature at mean density, $T_0$, as a function of redshift.
% Blue solid line shows the $512^3$ \arepo simulation, red dashed the $512^3$ \gadget simulation.
% The black dotted line shows the $256^3$ \arepo simulation, while the \gadget simulation 
% with $256^3$ particles is indistinguishable from the larger \gadget simulation.
% See Equation \ref{eq:meantemp} for the definition of $T_0$.
% }
% \label{fig:temp}
% \end{figure}
%%%%%%%%%%%%%%%%%%%%%%%%%%%%%%%%%%%%%%%%

For an analysis of the Lyman-$\alpha$ forest, we generated $16000$
simulated \Lya spectra with $1024$ pixels each from our simulations.
The positions of the spectra were chosen at random, and
particles/cells were interpolated to the lines of sight using an SPH
kernel. We verified that using cloud-in-cell interpolation for 
\arepo~did not affect our results. The
optical depth from the absorption due to each particle was calculated
as described in detail in \cite{Bird:2011}. In order to avoid
contaminating the spectra with strong absorbers, we did not apply a
self-shielding correction here. \cite{Bolton:2008} found evidence for
an inverted temperature-density relation in the \Lya forest, so that
lower density regions have a higher temperature.  Mechanisms proposed
for reproducing this include helium reionisation \citep{McQuinn:2009}
or volumetric heating from blazars \citep{Puchwein:2012}.  As our
purpose in this paper is a code comparison, we did not attempt to
model this in our simulations.
\edit{Thus, our model produces a temperature density relation 
$T \propto \rho^{\gamma-1}$ where $\gamma$ asymptotes towards $1/1.7$ 
at low redshift \citep{Gnedin:1997}, rather than the observed value of $\gamma \sim 1$.
At $z=3$ we have $\gamma = 1.55$.} 
 
We define the transmitted flux as $\mathcal{F} = \exp{(-\tau)}$, 
where $\tau$ is the optical depth. A
completely transparent Universe has $\mathcal{F} = 1$. Observations
have determined the effective optical depth, $\tau_\mathrm{eff} =
\overline{\mathcal{F}}$, where $\overline{\mathcal{F}}$ is the mean
transmitted flux, the one-dimensional flux PDF (a histogram of the
flux from each spectral pixel) and the flux power spectrum.
% The power spectrum is usually computed in velocity units; km/s, where $1\, \mathrm{s}/\mathrm{km} =  H(z)/(1+z) \hMpc$. 
We extracted all three of these quantities from our simulations and
compared the results of \arepo\ and \gadget.  We checked convergence using 
the simulations with $256^3$ particles. Figure
\ref{fig:tau_eff} shows the effect of a changing particle number on
$\tau_\mathrm{eff}$. \gadget~converged at the $2\%$ level for
$\tau_\mathrm{eff}$ and \arepo~to $\sim 4\%$, but this convergence
becomes significantly poorer for $z> 3.5$ in both
codes. \cite{Bolton:2009} found that this occurs because at high
redshift the \Lya transmission is dominated by progressively less
dense regions, which are less well resolved. 

%%%%%%%%%%%%%%%%%%%%%%%%%%%%%%%%%%%%%%%%
\begin{figure}
% \centering
\includegraphics[width=0.45\textwidth]{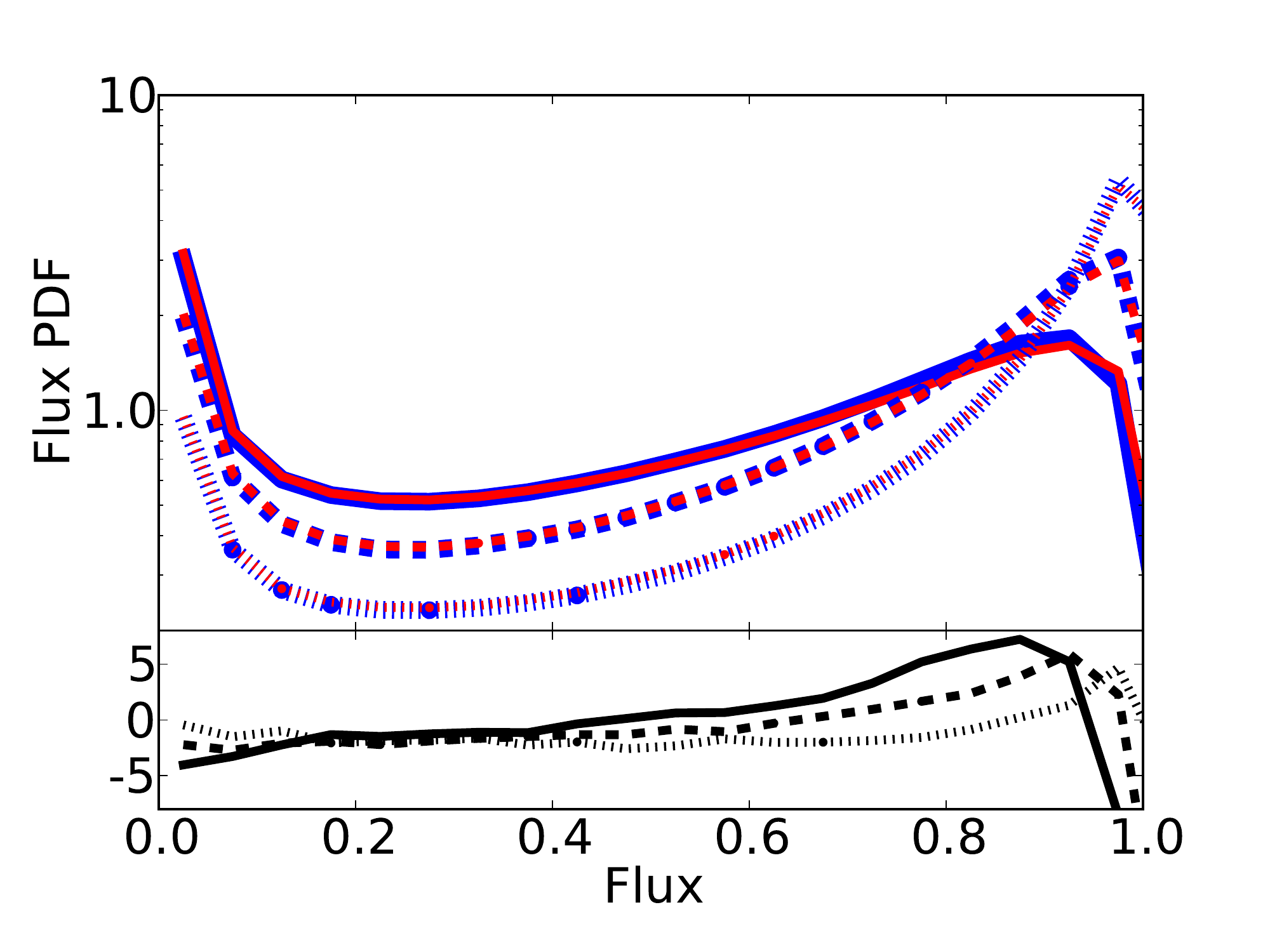}
\caption{The flux PDF for \arepo\ (thin red lines) and \gadget\ 
  (thick blue lines), at redshifts $z=3.0$ (solid), $z=2.6$ (dashed) and $z=2.0$
  (dotted).  Zero flux corresponds to total absorption. Lower panel shows
  the relative change in percent, with positive numbers corresponding
  to a larger PDF in \gadget.
}
\label{fig:fluxpdf}
\end{figure}
%%%%%%%%%%%%%%%%%%%%%%%%%%%%%%%%%%%%%%%%

%%%%%%%%%%%%%%%%%%%%%%%%%%%%%%%%%%%%%%%%
\begin{figure}
% \centering
\includegraphics[width=0.45\textwidth]{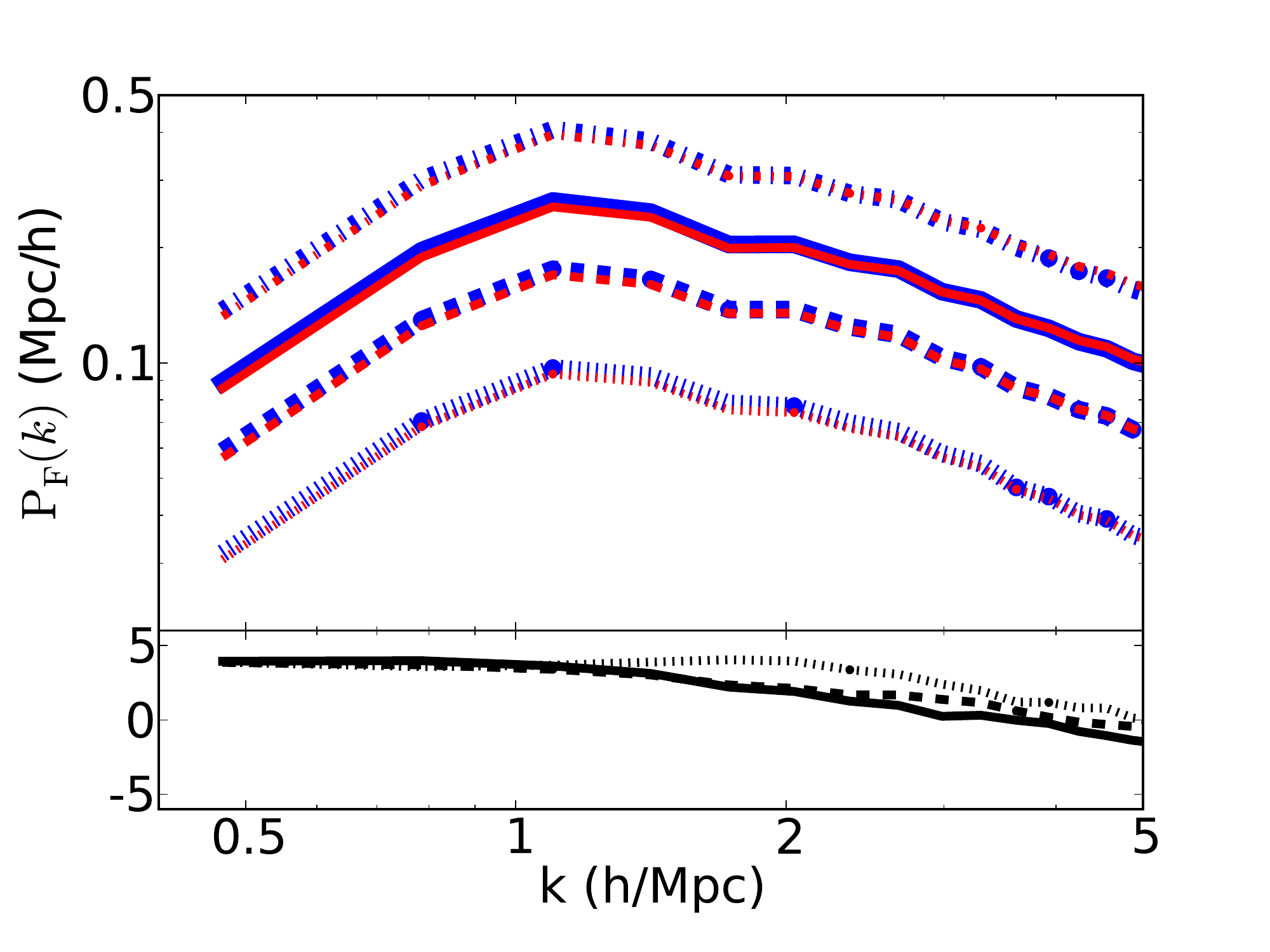}
\caption{The flux power spectrum for \arepo\ (thin red lines) and
  \gadget\ (thick blue lines), at redshifts $z=3.4$ (dot-dashed), $z=3.0$
  (solid), $z=2.6$ (dashed) and $z=2.0$ (dotted). The lower panel shows the
  percentage change, with positive numbers corresponding to more power
  in \gadget.  }
\label{fig:fluxpow}
\end{figure}
%%%%%%%%%%%%%%%%%%%%%%%%%%%%%%%%%%%%%%%%

In more detail, the effective optical depth is slightly lower in
\arepo, by $4\%$ at $z=4$, $2\%$ at $z=3$ and $1 \%$ at $z=2$.  Note
that $\tau_\mathrm{eff}$ is known observationally to $\sim 4\%$ at
$z=3$ \citep{Viel:2009}.  This small difference is due to a slight
increase in the mean temperature, $T_0$, of the \Lya absorbers,
defined to be gas with $T < 10^5$ K and $\rho < \rho_\mathrm{c}$,
where $\rho_\mathrm{c}$ is the critical density. $T_0$ in \arepo~is
$\sim 240\,{\rm K}$ ($2\%$) higher than in \gadget~at $z=2-4$. The
\Lya forest is assumed to be in ionisation equilibrium with the UVB,
so a higher temperature produces a greater ionisation fraction, thus
decreasing the effective optical depth.  These changes are somewhat
larger for the lower resolution simulations and thus are likely to be
reduced further with higher numerical resolution. 
It is the standard procedure, when comparing to data, to rescale 
simulated spectra to have the same mean flux as is observed. 
As we are performing a code comparison, we do not rescale our spectra for the presented results, but
we checked that it did not significantly affect our conclusions.

%%%%%%%%%%%%%%%%%%%%%%%%%%%%%%%%%%%%%%%
\begin{figure*}
% \centering
% \subfigure[\gadget]{
% \includegraphics[width=0.45\textwidth]{plots/phase_gad_512.pdf}
% }
% \subfigure[\arepo]{
\includegraphics[width=0.95\textwidth,trim=0 90 0 90,clip]{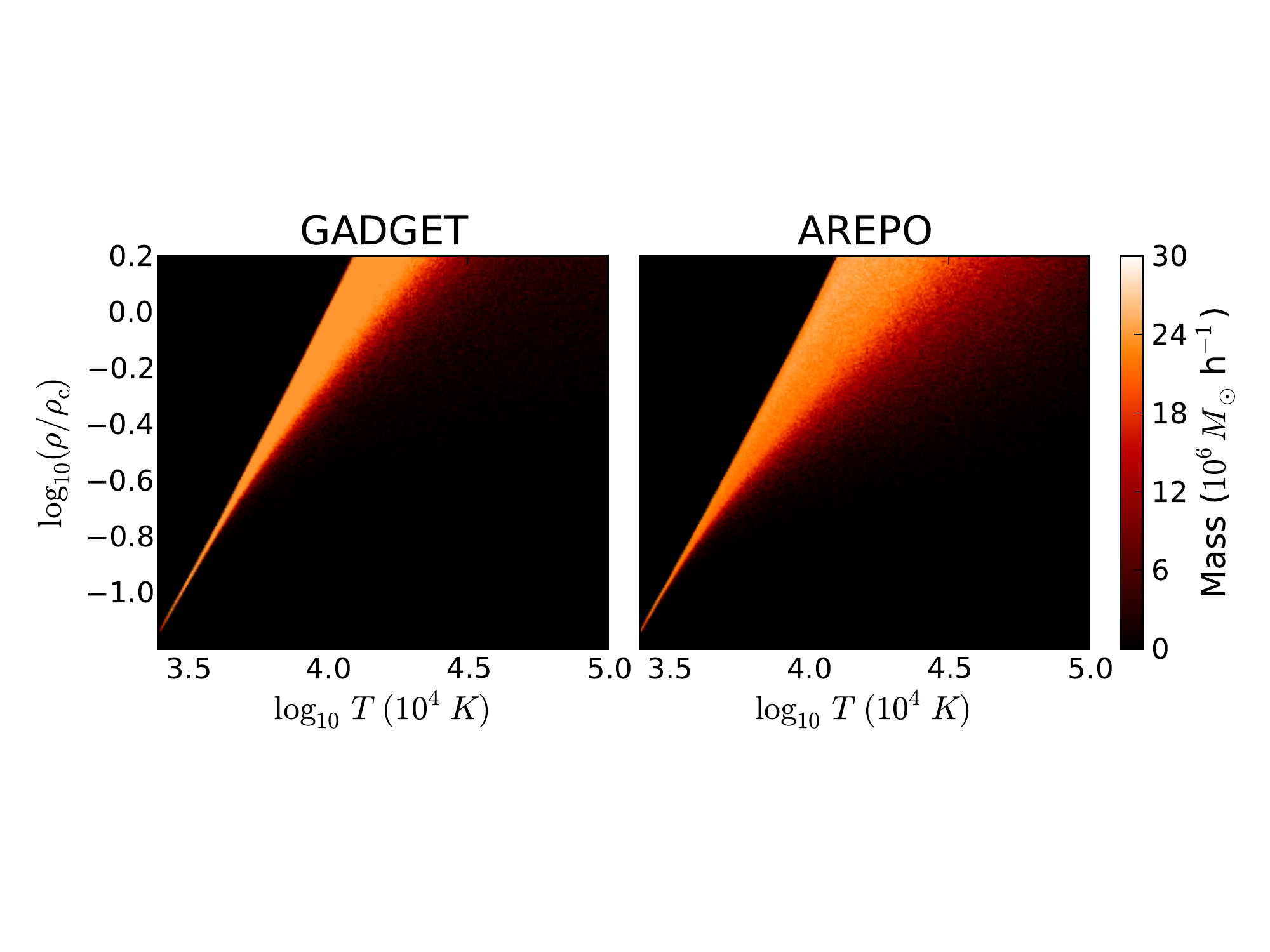}
% }
\caption{Two-dimensional mass-weighted histograms of the total gas
  mass as a function of temperature and density, at $z=3$, with
  $2\times 512^3$ resolution.  The colour scale shows the total mass
  of gas elements found in a given temperature-density bin.}
\label{fig:phase_512}
\end{figure*}
%%%%%%%%%%%%%%%%%%%%%%%%%%%%%%%%%%%%%%%

Figure~\ref{fig:fluxpdf} shows the flux PDF extracted from our
simulations in three redshift bins, chosen to match observations.
There are small differences between the codes; for $\mathcal{F} < 0.6$
the flux PDF is larger in \arepo, but for higher transmission regions
at $z=3$, the trend is reversed and \arepo~produces a lower flux PDF.
Figure \ref{fig:fluxpow} shows a similarly sized effect on the power
spectrum of the flux; \arepo~predicts a slightly lower power
spectrum. For comparison, the typical uncertainty in the observed flux
PDF is $5-7\%$ and $5-10\%$ in the flux power spectrum. These
are of the same order as the differences we find here between the
numerical codes.  We attribute these changes in the flux PDF and flux
power primarily to subtle shifts in the temperature-density
distribution. Figure~\ref{fig:phase_512} shows a mass-weighted
histogram of the temperature and density of gas elements.  We can see
that although the particles follow a similar temperature-density
relation overall, \arepo~produces a wider spread in temperatures for
gas near the critical density; this trend continues for higher density
gas until $\rho \sim 10 \rho_\mathrm{c}$.  

This difference was somewhat larger for lower-resolution simulations 
with $256^3$ particles, which might naively suggest that
\arepo~is converging more slowly in under-dense regions than
\gadget. However, it is not completely clear that the two codes are
converging to the same result, because the differences we see could
well be the effect of weak structure formation shocks. While these are 
followed accurately in \arepo, they are largely lost in
\gadget, due to SPH's poorer shock-resolving capability
\citep[see, e.g.][]{Keshet:2003}.
We would thus expect that a fully resolved \arepo~simulation would still produce
more gas elements scattered off the mean temperature density relation
than \gadget, consistent with our results. \edit{We checked, using a simulation with
a reduced Courant factor, that Figure~\ref{fig:phase_512} was not affected by 
the timestep.}

We recall that we have not included feedback in our simulations; this
is expected to affect the \Lya flux PDF at the $5\%$ level
\citep{Bolton:2008, Viel:2012} at $z=2-2.5$. Feedback processes such
as galactic winds may interact with the gas dynamics differently in
\arepo\ than in \gadget, potentially adding further systematic
differences between the codes of a similar magnitude.

\section{Conclusions}
\label{sec:conclusions}

We compared the statistics of neutral hydrogen absorption
in the SPH code \gadget~with that in the moving mesh code \arepo.
Our aim was to understand how known differences in the treatment of the
fluid equations manifest themselves as changes in the observable
properties of DLAs, LLS and the \Lya forest.  There were significant qualitative
differences: in \gadget, DLAs and LLS in large halos were primarily produced
by small spherical objects, and the only difference between the two
classes of absorbers was the central density. However, in the \arepo\ simulations,
DLAs and LLS came from quite different sources. The bulk of the DLA
cross-section in a large halo was from a central disc, while LLS were
produced in filamentary structures. This made little
quantitative difference to the DLA cross-section, but it suggests a 
different interpretation of high column density systems and could 
potentially be reflected kinematically in detailed line profile shapes.
We would suggest that any future studies sensitive to the detailed 
distribution of neutral hydrogen inside massive halos, especially LLS, 
should avoid using SPH.

We found that \gadget~produced clouds of clumpy substructure around 
halos with $M > 10^{11} \Msun$, which were essentially absent in \arepo.
These \gadget~clumps had a peak column density of $10^{18} - 10^{19}\,
{\rm cm}^{-2}$ and so boosted the LLS cross-section of these halos
significantly. There was a similar, but somewhat smaller, effect on the DLA
cross-section. Furthermore, halos in \arepo\ had central densities which were 
lower than their counterparts in \gadget. This significantly 
lowered the DLA cross-section in \arepo\ for objects sensitive to the density 
in the central $5 \kpch$ (in practice, halos with $M < 10^{10} \Msun$ at $z > 2$),
but did not greatly affect the LLS. 

These systematic changes made the halo mass -- cross-section 
relation shallower where it was dominated
by the changes to large halos and steeper where it was dominated by
changes to small halos. The former occurred for LLS, and DLAs at
redshift $z=2$, and the latter for DLAs at $z=3$ and $4$.
Furthermore, both changes acted to reduce the overall abundance of DLAs
in \arepo. The DLA abundance for \gadget~simulations is 
somewhat in excess of the observed value. This
discrepancy can be removed with the incorporation of galactic winds
into the simulation \citep{Nagamine:2004a, Tescari:2009}, but we found
that our \arepo~simulations already substantially reduce it, even
without strong feedback.  This was mostly due to the reduced
cross-section of small halos; large halos are sufficiently rare that
changes in their cross-section do not have a large effect on the total
abundance. It also shows that a calibration of feedback strength from 
the DLA abundance would be compromised by numerical effects in SPH.

The column density function shows a similar pattern; a modest reduction in
amplitude for $N_\mathrm{HI} = 10^{20} - 10^{22}\,{\rm cm}^{-2}$ and
$z > 2$, driven by the lower central density of small halos.  
We found that the amplitude of the column
density function for $N_\mathrm{HI} > 10^{21}$ cm$^{-2}$ is still much
larger than observed. \edit{This discrepancy is much larger than the differences 
between our two simulations}, which can be interpreted as evidence that some
feedback in the form of outflows is needed.
Although the column density function for large halos was significantly 
reduced in \arepo~by the changes in their substructure, the relative 
rarity of these halos meant that they did not have a strong impact on 
the total column density function.
Note, however, that feedback processes reduce the amplitude of the high-end column density function
by removing preferentially the high-column density cells in small
halos; the more massive the halo, the less it is affected by winds
\citep{Tescari:2009}. Thus, although it appears as if the reduced
substructure around large halos only produces subtle changes in
this statistic, once galactic winds have blown the small halos away, 
what remains may be very greatly affected.

Finally, we examined basic \Lya forest statistics and found
that differences between the codes were typically at the few percent level.
This was due partly to a slightly changed thermal history (which is
typically marginalised out in cosmological studies of the \Lya
forest) and partly to an increased width in the
temperature-density relation of the gas in \arepo. 
These differences are small when compared to 
changes in the high column density systems. 
This is not unexpected; the evolution of the \Lya forest is dominated 
by gravitational effects, hence issues such
as the accuracy problems of SPH for fluid instabilities do not play an
important role \citep[a similar result was found by][]{Regan:2007}.
They are however comparable to the statistical uncertainties of current data, and 
thus may bias derived parameters. Although a full cosmological analysis 
is beyond the scope of this paper, it seems likely
that the largest difference will occur in the derived thermal 
history of the IGM, since the change in the flux power spectrum 
came predominantly from an alteration in the temperature-density relation. 
These effects will also have to be taken into account 
when analysing upcoming \Lya experiments such as 
BOSS \citep{Slosar:2011}, which are expected to have statistical errors 
an order of magnitude smaller than those of current data.

There are many questions about high column density systems which
remain unanswered by the present work. Although our \arepo~simulations
reduce the discrepancy between DLA simulations and observations,
completely eliminating it still seems to require stronger feedback
processes.  Furthermore, SPH simulations have historically failed to
reproduce the velocity width of metal lines in DLAs, due to the
suppression of mixing in SPH, which leaves metals concentrated in
small clumps, instead of being spread out through the halo
\citep{Tescari:2009}. Here the new \arepo\ code offers a lot of
potential for progress. We intend to study these questions in future
work, based on a more comprehensive model for feedback and metal
enrichment (Vogelsberger et al., 2013, in preparation).

\section*{Acknowledgements}

SB would like to thank Yajima Hidenobu for sending us his collation of
observational data on the column density function.  SB is supported by the
National Science Foundation grant number AST-0907969 and the Institute
for Advanced Study. DS acknowledges NASA Hubble Fellowship through grant
HST-HF-51282.01-A. MZ is supported by the National Science Foundation 
under PHY-0855425, AST-0907969 and PHY-1213563 and by the David and Lucile Packard Foundation. 
VS acknowledges financial support from the Klaus Tschira
Foundation and the Deutsche Forschungsgemeinschaft  through SFB 881, ``The
Milky Way System''. LH is supported by NASA ATP Award NNX12AC67G.

%%%%%%%%%%%%%%%%%%%%%%%%%%%%%%%%%%%%%%%%%%%%%%%%%%%%%%%%%%%%%%%%%%%%%%%%%%%%%%
\bibliography{comparison}
\label{lastpage}
\end{document}